\documentclass[eqsecnum,preprint]{aastex}
\usepackage{natbib}
\usepackage{epsf}
\begin{document}
\bibliographystyle{plain}
\title{The Dynamics of the Merging Galaxy Cluster System A2256: Evidence for
a New Subcluster}
\author{Robert C. Berrington}
\affil{ASEE Posdoctoral Fellow, Naval Research Laboratory, Code 7653, 
Washington, DC 20375-5352\\
Electronic Mail:  rberring@gamma.nrl.navy.mil}
\author{Phyllis M. Lugger and Haldan N. Cohn}
\affil{Department of Astronomy, Indiana University, Bloomington, IN 47405;\\
Electronic Mail:   lugger@astro.indiana.edu, cohn@astro.indiana.edu}

\def\ea{{\it et al.}}
\def\etal{{\it et al.}}
\def\mev{\mbox{$\;{\rm MeV}$}}
\def\kev{\mbox{$\;{\rm keV}$}}
\def\gev{\mbox{$\;{\rm GeV}$}}
\def\ergs{\mbox{$\;{\rm ergs}$}}
\def\ergss{\mbox{$\;{\rm ergs\ s^{-1}}$}}
\newcommand{\kmsec}{\mbox{$\rm{km\ s^{-1}}$}} 
\newcommand{\msun}{\mbox{$M_{\odot}$}}  
\newcommand{\KC}{\mbox{\small{$_{\rm KC}$}}} 
\def\gtorder{\mathrel{\raise.3ex\hbox{$>$}\mkern-14mu
             \lower0.8ex\hbox{$\sim$}}}
\def\ltorder{\mathrel{\raise.3ex\hbox{$<$}\mkern-14mu
             \lower0.8ex\hbox{$\sim$}}}

\begin{abstract}
We present 236 new radial velocities of galaxies in the cluster A2256
measured with the WIYN Hydra multi-object spectrograph.  Combined with
the previous work of \citet{FAB89}, we have velocities for a total of
319 galaxies of which 277 are cluster members.  In addition to the new
radial velocities, we present a $3 \times 3$ image mosaic in the R
band of the central $19\arcmin \times 19\arcmin$ region of A2256, from
which we obtained photometry for 861 galaxies.  These data provide
strong evidence for a merger event between two groups.  In addition,
we present evidence for the presence of a third group, on the outer
reaches of the system, that is just now beginning to merge with the
system.
\end{abstract}


\section{Introduction}
\label{sec:introduction_A2256}

Originally thought to be a highly relaxed or evolved cluster, A2256 was shown
by \citet{DRESSLER78} to show evidence for equipartition in energy.  With a
richness class 2, A2256 is surpassed only by 5\% of the clusters found in
Abell's (1958) catalog.  \citet{FABER77} measured the velocity dispersion to
be $1351\ \kmsec$, making it one of the highest velocity dispersions of known
galaxy clusters.  \citet{CARTER80} showed the galaxy distribution to be
elliptical.  \citet{FAB89} increased the number of measured radial velocities
to 89 and found a flat velocity histogram.  They first suggested the presence
of substructure, and interpreted this as indicating that A2256 represents a
merger in progress.

In the X-ray, A2256 is a luminous and hot cluster with a central temperature
of $\sim\!\!7$\kev.  First imaged by the {\em Einstein} IPC, the X-ray
emission from A2256 was found to depart significantly from spherical symmetry
\citep{FAB84}.  Observations by the R\"{o}entgen Satellite (ROSAT) resolved
the non-spherical X-ray emission into two maxima in the central region of the
cluster \citep{BRIEL91}. Briel and Henry (1994) used several Position
Sensitive Proportional Counter (PSPC) observations with the ROSAT Observatory
to obtain a temperature map of A2256.  They found an extremely complicated
temperature structure with two opposing 12.0 keV lobes almost perpendicular to
a proposed infall direction for the two merging subgroups.  However, the
uncertainties of this temperature map are large.  Later observations by the
Advanced Satellite for Cosmology and Astrophysics (ASCA) did not observe a
similar central temperature structure, but did find a steep radial temperature
profile in agreement with ROSAT data \citep{MARKEVITCH96}.

A2256 also exhibits a number of striking features in the radio regime, as
discussed by \citet{ROTTGERING94}.  It has at least four galaxies with
head-tail morphologies, placing it amongst three to four other clusters with
comparable numbers of confirmed head-tail radio features.  It also contains
one of the longest and narrowest head-tail structures at 700 kpc in length and
less than 2.5 kpc in width.  A Z-shaped structure with an ultra-steep spectrum
has been tentatively identified with a cluster galaxy, but it eludes
definitive classification.  The most pronounced radio features in A2256 are
two extended halos with spatial dimensions that are estimated at 1.0 Mpc by
0.3 Mpc.

A2256 is well suited for the study of substructure and the role of dominant
galaxies.  Its optical structure is very much like that of the Coma cluster
(A1655) with a central pair of dominant galaxies, each at the center of a
concentration of fainter galaxies.  The galaxy distribution also indicates the
likely presence of the substructure that is seen in the X-ray images.  In this
paper, we present new optical imaging of the central region of the cluster and
278 new radial velocities obtained with the WIYN telescope.

In section \ref{sec:data}, we describe both new imaging and spectroscopy
obtained by the Wisconsin-Indiana-Yale-NOAO (WIYN) telescope.  We also include
the results of the data analysis.  In section \ref{sec:analysis}, we review
the statistical methods for searching for substructure.  Section
\ref{sec:discussion} includes a discussion of the implications of the
observations and the statistical tests.  In section
\ref{sec:summary_and_conclusion}, we summarize the results of this study.

\section{Data}
\label{sec:data}

\subsection{CCD Imaging Observations}
\label{sub:ccd_imaging_observations}

Since the previous study of the galaxy distribution in A2256 has been based on
scans of Palomar Observatory Sky Survey plates \citep{FAB89}, we obtained deep
CCD images of the central region of A2256.  We used the WIYN 3.5\,m telescope
at the Kitt Peak National Observatory (KPNO) to acquire total of 18 images in
the form of a $3\times3$ mosaic centered on the cluster's center of $\alpha =
17^{h}\,4^{m}\, 1\fs1,\,\delta = 78^{\circ}\,38\arcmin\,3\farcs6$ (J2000.0).
At the time of these observations, the WIYN imager used a
$2\rm{k}\times2\rm{k}$ Tektronix CCD to image a $6\farcm7$ square field at a
scale of $0\farcs20$ per pixel.

We obtained two 600 s Kron-Cousins R--band [R\KC] exposures per field.
The fields are labeled east to west by C, D, and E, and are labeled
north to south by 3, 4, and 5.  The central field D4 of the mosaic is
centered on the cluster center.  The surrounding fields were offset by
steps of $6\farcm2$ in RA and Dec., resulting in a 100 pixel overlap
in fields.  However, a small portion of the sky was missed between
images D5 and E5 due to an error in telescope pointing.  Otherwise,
coverage was complete over all the mosaic.  Table 1 shows the RA and
Dec.\ of the center of each image frame, as well as the observation
date.

\placetable{FIELDS}

We used the imred.ccdred package in IRAF\footnote{IRAF is distributed
by the National Optical Astronomy Observatories, which are operated by
the Association of Universities for Research in Astronomy, Inc., under
cooperative agreement with the National Science Foundation.} to reduce
the images.  All images were bias-subtracted, flat-fielded, and
dark-current subtracted.  In addition, each frame contains a dead
column and a hot column that were removed using the fixpix procedure.
Each frame pair was coadded by the imcombine procedure after being
matched, translated and rotated by the geomap/geotran procedures.
Most images required a shift of less than a pixel.  However a few
images were taken with an appreciable difference in time between the
two pair members, and thus required shifts on the order of 50--100
pixels.

To find galaxy positions, we used the Faint-Object Classification
Analysis System (FOCAS) to locate and produce positions for all the
galaxies within each frame.  To perform the astrometry, we obtained a
list of approximately 300 stars within the $3\times3$ WIYN mosaic from
the Digital Sky Survey (DSS) Quick-V scans.  Any star found to be
saturated on a WIYN image was removed from the list.  The unsaturated
stars were centroided on each WIYN frame and used to find a plate
solution, based on at least 20 stars.

We used FOCAS to perform the photometry on each WIYN frame.  Instrumental
magnitudes were obtained from all the fields, and transformed into the
R\KC--band.  To correct for galactic extinction, we adopt the following
relation \[A_{\rm{R\KC}} = 2.47\ E(B-V)\] which was obtained from the
transformations of \citet{FERNIE83}.  Values of E(B-V) were obtained from the
color excess maps of \citet{BURSTEIN82}.  K-corrections conform to the
procedures described by \citet{POSTMAN95} and \citet{PENCE76}.  In Table 3, we
report all the R\KC--band magnitudes obtained from the WIYN images.

From the $3\times3$ WIYN image mosaic, we obtained a catalog of 861 galaxies.
All matches from overlapping CCD frames were removed to prevent any
redundancy.  Because the astrometry was done using a different set of stars
for each frame, all the matches were extracted and their positions were
checked against each other to estimate the accuracy of the astrometry.  A
total of 30 matching pairs were found and the difference in positions were
calculated.  Astrometry of the matching pairs is susceptible to a number of
errors.  The plate solution is worst at the edge of the plate since the
distribution of the stars used in finding the plate solution may not fully
cover the CCD from edge to edge.  Because galaxies are extended objects,
portions of galaxies near the edge of a WIYN frame are sometimes missing from
the image.  For these galaxies, the image that contained the entire galaxy was
used for the both the photometric and the astrometric work.  By comparing the
matching pairs of galaxies on adjacent WIYN frames, we estimate accuracy of
the galaxy positions to be $0\farcs3$.  

\subsection{Spectroscopy}
\label{sub:spectroscopy}

We report radial velocity measurements for 278 galaxies, of which 48 are
common with the samples measured by Fabricant \ea\ (1989) and \citet{FABER77}.
By combining these new measurements with velocities for 41 additional galaxies
from these two previous studies, we have developed a sample of 319 galaxy
velocities for our analysis.  All new radial velocities were obtained using
the Hydra Multi-Object Spectrograph on the WIYN 3.5\,m telescope at KPNO\@.
Both the imaging data and the spectroscopic data were taken over several runs.
An observation log for the spectroscopic data is given in Table
\ref{HYDRASETUP}.  

In addition to the 861-galaxy catalog obtained from the WIYN $3\times3$
image mosaic, we obtained an image from the DSS Quick-V scans.  Using the
FOCAS package, we obtained a list of galaxies within $\pm 0.5^{\circ}$ of
the cluster center.  The combination of this list with the catalog from the
WIYN $3\times3$ image mosaic and the galaxy list from Fabricant \ea\ (1989)
produced a total list of 1090 galaxies, which was used in the galaxy
selection.

\placetable{HYDRASETUP}

The Hydra spectrograph contains 250 ports within which are placed fibers that
act as apertures for the spectrograph.  Either of two 100-fiber cables sets
may be selected, one optimized for the blue and one for the red.  Galaxy
selection was carried out using the whydra program which optimizes fiber
placement based on a set of user defined weights, subject to a set of fiber
placement rules.  Our method assigned weights according to the magnitude of
the galaxy and required that at least 20 fibers be placed on blank sky.

For the June 1996 run, we chose the 500--1000 nm band and the red cable to
observe the H$\alpha$, H$\beta$, Mg b, and Na D lines.  The red fiber set uses
a $2\arcsec$ size fiber that is optimized for the $450-1,\!000$ nm wavelength
range.  Due to the presence of strong sky lines, the spectra redward of
$\sim\!\!700$ nm were not usable for velocity measurements.  Exposure sets of
$3\times30$ min were used for the first two fields and $2\times20$ min was
used for the third field.  For the June 1997 and June 1998 runs, we chose a
higher dispersion and the blue cable to concentrate on the 430--720 nm region.
The blue fiber set uses a $3\arcsec$ size fiber.  The exposure sets with the
blue cable were all $2\times1$ hour.  We bracketed each set for both runs by a
Th-Ar comparison lamp exposure for wavelength calibration.  Each exposure was
bias subtracted using the IRAF ccdred package.  All spectra were flat fielded,
dispersion corrected and extracted using the IRAF hydra package.

A minimum of 20 fibers were used to obtain sky spectra, which were
combined to form a sky spectrum template.  All galaxy spectra were then sky
subtracted.  The only problematic sky feature was the $\lambda\lambda$557.7 nm
[O I] line.  Although it usually subtracted cleanly, it was sometimes
necessary to remove it manually.  All exposures for each setup were averaged
together using a $3\sigma$ cosmic-ray clip.  Figure \ref{SPECTRA} shows a
typical spectrum with target lines marked.

We extracted radial velocities using the IRAF fxcor package.  For both
the red and blue cable setups, only the 450--720 nm spectral region was
used in determining the radial velocities.  Three templates of objects
with known radial velocities were obtained for the cross correlation.
Two stellar templates, AGK224 and HD 223094, and one galaxy template
NGC 7331, of radial velocities $-49.5$, 26.0, and 819.0 \kmsec,
respectively, were each used to obtain the heliocentric velocity of
each galaxy.

\placefigure{SPECTRA}

Depending on the spectrograph setup and galaxy magnitude, the uncertainties of
the radial velocities varied over the range of $10-120$ \kmsec\ with a median
uncertainty of 52 \kmsec.  In the June 1996 run, a total of 31 galaxies done
by Fabricant \ea\ (1989) were redone.  Figure \ref{DELTACZ} shows a comparison
of our measured velocities with those of Fabricant \ea\ (1989) as a function of
$m_{r}$.  The mean offset is $-55$~\kmsec, with a standard deviation of
98~\kmsec\ and and an error of the mean of 18~\kmsec.  While the mean offset
is somewhat larger than expected, it is still small compared with
the combined measurement errors for the two studies.  

\placefigure{DELTACZ}

In addition to the velocities redone in the June 1996 run, 45 galaxies from
Fabricant \ea\ (1989) were redone in the June 1997 and June 1998 runs.  We
performed a comparison of these galaxy velocities with both those that were
measured in June 1996 and those of Fabricant \ea\ (1989).  We found all the
velocities measured in June 1997 and June 1996 to be consistent with each
other within the $1\sigma$ level.  The offset between the values measured by
Fabricant \ea\ (1989) and our values persisted in the June 1997 and June
1998 measurements, with a mean offset of $-70$~\kmsec, a standard deviation
of 84~\kmsec, and an error of the mean of 13~\kmsec.  Figure
\ref{BERRDELTACZ} shows a comparison between the Fabricant \ea\ (1989) and
the June 1996 run versus the magnitude $m_{r}$.  We checked the wavelength
of the $\lambda\lambda$ 557.7~nm sky line to check for any instrument
dependent offsets.  No such offsets were found in any of the hydra setups.
Because our two spectrograph setups used not only different fibers, but also
different central wavelengths and gratings, we conclude that the offset
between our values and those of Fabricant \ea\ (1989) most likely reflects
the lower accuracy of the previous study.  In any case, the offset is within
the measurement errors.

\placefigure{BERRDELTACZ}

All radial velocity data obtained in the WIYN observing runs described
in Table \ref{HYDRASETUP} are reported in Table \ref{CATALOG}.  For
each velocity reported by the IRAF fxcor package, a visual check
between the velocity standard and the cluster galaxy was performed to
avoid accidental selection of any spurious peaks in the cross
correlation.  During the three nights a total of five different setups
were used for A2256.  In each setup, at least 20 fibers were used to
obtain the sky spectra, and on average 11 fibers were used for the
previously measured galaxies.  The remaining $\sim50$ fibers were used
for galaxies selected from the DSS Quick-V scans.

\placefigure{COMPLETE}

Figure \ref{COMPLETE} shows the fraction of galaxies with radial
velocity measurements versus R\KC\ magnitude for the combined data
set.  All magnitudes were converted to the R\KC\ magnitude system for
comparison.  Galaxies without magnitude information were not used in
the determination of the fraction of galaxies with radial velocities.

\subsection{Catalog}
\label{sub:catalog}

We present the results of our galaxy spectroscopy in Table
\ref{CATALOG}\footnote{Table \ref{CATALOG} contains only the galaxies with
measured redshifts.  The entire galaxy catalog of 1090 galaxies that combines
those galaxies obtained from Fabricant \ea\ (1989) and the WIYN $3\times3$
mosaic will be made available electronically from the Astronomical Data Center
(ADC).  The ADC's internet site hosts WWW and FTP access to the ADC's archives
at the URL http://www.gsfc.nasa.gov}.  Column 1 contains a galaxy
identification number that will be used throughout this paper.  Columns 2 and
3 give the equatorial coordinates of each galaxy in the J2000.0 epoch.  All
galaxies are sorted by increasing RA\@.  Column 4 contains the total R\KC-band
magnitude of the galaxy as reported by FOCAS.  The previous studies of
\citet{FABER77} and Fabricant \ea\ (1989) provide isophotal magnitudes.  We
first compared the Fabricant \ea\ (1989) r magnitudes with our R\KC\
magnitudes and found a systematic difference that is similar to what they
noted when comparing their magnitudes with the F magnitudes reported by
\citet{FABER77}.  Column 5 reports the $1\sigma$ error on the R\KC\ magnitudes
reported in column 5.  Column 6 presents the radial velocities obtained either
in the present study or in the previous studies of \citet{FABER77} and
Fabricant \ea\ (1989).  The new velocity measurement is reported for those
galaxies that were remeasured.  Column 7 reports the $1\sigma$ velocity errors
of the velocities in column 5.  As indicated in column 5, these errors are
either as reported from the IRAF fxcor package or the errors reported in the
previous studies.  All cross identifications between Fabricant \ea\ (1989) and
the current work are indicated in column 8.

\placetable{CATALOG}

\section{Analysis}
\label{sec:analysis}

\subsection{The Search for Substructure}
\label{sub:search_for_substructure}

If a galaxy cluster is a dynamically relaxed system, then the spatial
distribution of galaxies should be approximately spherical and the velocity
distribution should be approximately Gaussian.  The presence of substructure
indicates a departure from this quasi-equilibrium state.  Substructure is
indicated by at least one of the following: (1) significant multiple peaks
in the galaxy position distribution, (2) significant departures from a
single Gaussian velocity distribution, and (3) correlated deviations from
the global velocity and position distributions.  This section presents the
results of a battery of statistical tests that we applied to search for such
substructure.  Table \ref{TESTS} provides a brief description of each test
that we applied to the data, along with references to more thorough
descriptions.

\placetable{TESTS}

Following \citet{YAHIL77}, we performed a $3\sigma$-clip on the velocity
sample to filter out likely foreground and background galaxies.  An
alternative technique to filter out the foreground and background galaxies,
based on gaps in the velocity histogram, has been proposed by Zabludoff \ea\
(1993).  We chose the $3\sigma$-clip technique because the sample returned
from the alternative technique differed by at most two galaxies at the high
redshift end of the velocity histogram.  All the statistical tests were run on
both the $3\sigma$-clip data set and a $5\sigma$-clip data set.  The results
were unaffected by the different populations.  The $3\sigma$-clip reduced the
sample of 319 galaxies to 277 members.  The galaxies removed from the sample
consisted largely of galaxies from a background group centered on the
approximate sky coordinates RA: $17^{h}\,07^{m}$, Dec: $78^{\circ}\,22\arcmin$
(J$2000.0$) and velocity $52,\!413.0$\ \kmsec.

Each test listed in Table \ref{TESTS} requires a normalization procedure.
For the one-dimensional tests, all calculations were normalized with respect
to 10,000 bootstrapped intervals.  The two-dimensional tests required an
azimuthal shuffling of the data; $10,\!000$ shuffles were used.  For the
three-dimensional tests, the velocities were randomized with respect to the
positions.  For all the three-dimensional tests, with the exception of the
Lee-3D test, 10,000 randomizations were used.  Due to the CPU cost, the
Lee-3D test was applied with only 1,000 shuffles.  To calculate the
confidence intervals of the tests, the value of the test statistic obtained
from the original data was compared to the distribution of values from the
the shuffled data.  The number of instances where the value of the test
statistic was greater for the shuffled data than for the original data was
divided by the total number of shufflings.  This value is the confidence
interval of the test.  This bootstrap procedure is followed because the
a priori probability distribution for the test statistic is often not
known.  

In Figure \ref{KMM1}, we present the velocity histogram of the 277
galaxies which are likely cluster members, together with a single
Gaussian fit.  While a quick visual inspection may suggest that the
single Gaussian provides a reasonable representation of the
distribution, a more detailed examination indicates several striking
deviations from Gaussian behavior.  The detailed statistical tests
that we have applied support the conclusion that the structure is in
fact non-Gaussian.  At high velocities, there is a striking group of
galaxies centered about $\sim\!\!19,\!600\ \kmsec$, and a relative
deficit between $19,\!000$ -- $19,\!200$\ \kmsec.  The remaining
histogram has a strong peak, located at $17,\!900\ \kmsec$, and
between $15,\!000$ -- $19,\!000$\ \kmsec appears highly skewed to the
lower velocity end of the distribution at $17,\!900$ \kmsec.
Furthermore there is a strong correlation between velocity space and
sky positional space that is unexpected from a relaxed system.

All one-dimensional tests performed are from the ROSTAT package
\citep{BEERS90}.  The results are presented in Table 8.  Column 1 gives the
name of the statistic.  Column 2 gives the value of the statistic, and column
3 indicates the probability that this value can be obtained given the null
hypothesis of no substructure.  Low probability values provide significant
evidence for substructure.  In the case of the one-dimensional tests, the null
hypothesis is that the velocity distribution is a single-component Gaussian.
In Table \ref{ROSTAT}, the u-test, W-test, $b_2$-test, $\sqrt{b_1}b_2$-Omnibus
test, and marginally the a-test are the only ones that show significant
deviations from a Gaussian in the $3\sigma$-clipped data set.

\placetable{ROSTAT}

A2256 has a nearly azimuthally symmetric galaxy distribution.  This makes it
difficult to detect substructure with two-dimensional tests, which are
applied to the projected galaxy distribution.  As a consequence, 2-D tests
were not stressed in our search for substructure.  The results of the 2-D
tests are presented in Table \ref{2dTESTS}.  In addition, it should be noted
that tests that are sensitive to elongations or ellipticity, such as the
Fourier-Elongation test, do not necessarily detect the presence of
substructure.  Two of the three two-dimensional tests showed evidence for
substructure.  These tests are the Lee-2D $l_{max}$ and the Lee-2D $l_{rat}$
test.  The results of the test indicate an elongation along a line which is
oriented at a position angle of $120^{\circ}$.  This direction is
approximately the same as the apparent merging axis of the cluster.

\placetable{2dTESTS}

With the large number of new velocities from our WIYN-Hydra spectra and the
more accurate positions obtained from the WIYN images, it is natural to
apply three-dimensional tests to the data set.  Of all the tests, the
three-dimensional tests were the most CPU intensive, with the Lee-3D test at
the top and the $\alpha$-test as the least computationally expensive test.
Typical run times for the calculations ranged from 10 minutes for the
$\alpha$-test on a 99 MHz HP-735 workstation to 50 hours for the Lee-3D test
on a 233 MHz Pentium II Linux system.  The results of the tests are
presented in Table \ref{3DTESTS}.

A number of modifications have been made to the tests.  Most importantly,
the tests have been altered to use bi-weight estimators in place of
classical estimators of the mean and standard deviations.  In addition, the
definition of a local sample has been changed to the $N_{local} = \sqrt{N}$
nearest neighbors as suggested by \citet{BIRD94}, instead of the canonical 11
nearest neighbors.  In our case $N_{local} = 14$.

\placetable{3DTESTS}

As can be seen in Table \ref{3DTESTS}, two of the four three-dimensional,
The Dressler--Shectman $\Delta$-test and the $\alpha$-test, provide evidence
for substructure at a $\gtorder 94\%$ confidence level.  Figure \ref{DELTA}
displays the Dressler--Shectman bubble plot for A2256.  It is immediately
apparent that there is a grouping of galaxies, to the north of the center,
that deviate significantly from the global values.  The galaxies that
correspond to large circles in the bubble plot comprise the small clump of
galaxies in the velocity histogram (Figure \ref{KMM1}) near $19,\!900$
\kmsec.  The Lee-3D test $l_{max}$ did not show evidence for substructure.
Since this test is designed for a bimodal distribution, it may be affected
by the presence of a multi-modal distribution.

\placefigure{DELTA}

The tests that show evidence for substructure are the u-test, the $b_2$-test,
the $\sqrt{b_1}b_2$-Omnibus test, the W-test, the Lee-2D $l_{max}$ and
$l_{rat}$ tests, the Dressler--Shectman $\Delta$-test, and marginally the
a-test.  Each test is sensitive to a different departure from the null
hypothesis of a unimodal Gaussian velocity distribution that is uncorrelated
with an azimuthally symmetric galaxy distribution.  Thus, it is not expected
that all tests will reject this null hypothesis.  The fact that six tests do
reject the null hypothesis at a confidence of $\gtrsim 95\%$ provides strong
evidence for the presence of substructure.

\subsection{Mixture Modeling}
\label{sub:mixture_modeling}

As indicated in section \ref{sub:search_for_substructure}, there is strong
evidence for the presence of substructure in A2256.  To quantify the nature of
substructure, we employed the KMM software package \citep{MCLACHLAN88} which
was previously used for this application by \citet{ASHMAN94}.  The KMM
software fits a user-specified number of Gaussians to a multi-variate
distribution, and provides a comparison of the quality of the multi-Gaussian
fit relative to a single-Gaussian fit.  This software is based on the
Expectation-Maximization (EM) algorithm which is described by
\citet{MCLACHLAN97}.  Each element in the distribution is then assigned to one
of the Gaussians based on maximum likelihood.

Figure \ref{KMM1} shows the velocity distribution histogram, along with the
best single-Gaussian fit.  At first glance, the single-Gaussian model may
appear to provide a reasonable fit, but the deviations are suspiciously
large.  To test the single-Gaussian fit, we applied the Double-Root-Residual
(DRR) test described by \citet{GEBHARDT91}.  The DRR test showed that the
peak located at $18,\!000$ \kmsec and the peak near $19,\!600$ \kmsec are
indeed significant deviations from the single-Gaussian fit.

To determine which multi-Gaussian model works best, we applied the KMM
software to the velocity and position data set for models with 2, 3, and 4
Gaussian components.  Thus, we have adopted a Gaussian model for the spatial
distribution within each subclump as well as for the velocity distribution.
The Gaussian model provides a convenient tool for deconvolving multiple
spatial components, rather than an accurate model for the spatial structure
of each subclump.

\placefigure{KMM1}

The 3-Gaussian model provided the best fit and the result is shown in Figure
\ref{3GAU}.  The mean, dispersion, and number of galaxies assigned to each of
the Gaussians are given in Table \ref{KMM3T}.  The improved fit of the the
3-Gaussian model, relative to a single-Gaussian model, is significant at the
99\% level.  In Figure \ref{3GAU}, it can be seen that Gaussian 3 corresponds
well to the high-velocity clump centered at $\sim\!\!19,\!600\ \kmsec$.

\placefigure{3GAU}

\placetable{KMM3T}

We also investigated whether the large velocity clump marked by Gaussians 1
and 2 has significant substructure.  To test for this we removed the galaxies
associated with Gaussian 3 and ran the same set of statistical tests on the
1--2 clump data set.  All 1-dimensional tests but the I-test showed
significant deviations from a Gaussian distribution.  

It is useful to observe the projected spatial distribution of the galaxies
belonging to each subclump.  This is shown in Figure \ref{A2256radec4}, where
the symbol type plotted for each galaxy indicates its subclump membership.
The localization of group 3 is immediately obvious.  In contrast, the 1 and 2
groups have a large overlap.  We interpret this as indicating that the 3
subclump is a small group in the process merging with the 1--2 system.  

\section{Discussion}
\label{sec:discussion}

\subsection{Cluster Dynamics}
\label{sec:cluster_dynamics}

In \S\ref{sub:mixture_modeling} we presented evidence for the presence of
three subclumps.  Previous analyses, based on a smaller velocity sample, have
detected only the 1 and 2 subclumps \citep{FAB89,BRIEL91,ROETTIGER93}.  To
check to see which systems are gravitationally bound, we applied the projected
mass estimator as described by \citet{HEISLER85}.  The masses are less than
the half the virial masses of the corresponding subclumps, potentially
indicating that the system is not in virial equilibrium.  In Table \ref{KMM3T}
the mass for each of the subclumps is given.

\placefigure{A2256radec4}

To perform the gravitational bound check we can rewrite the Newtonian energy
criterion for the two-body problem as:
\begin{equation}
V_{r}^{2} R_{p} \leq 2 G M \sin^{2} \alpha \cos \alpha \label{bound},
\end{equation}
where $V_{r}$ is the line-of-sight velocity, $R_{p}$ is the projected
separation of the clump centers, $\alpha$ is the projection angle measured
from the sky plane, and $M$ is the total mass of the system \citep{BEERS82}.
Since equation (\ref{bound}) is based on a two-body analysis, we applied it
to subclumps 1 and 3 in the following manner.  For clump 1, we assumed that
the mass of clump 3 is negligible, and the two point masses are subclumps 1
and 2.  For group 3, we assumed that subclumps 1 and 2 are bound and
constitute the other point mass.  If all these systems are bound, they are
likely to be interacting with each other.  As a consequence, the
first-ranked galaxy (FRG) may not be located at the center of a local
potential well.  To minimize the effect of any FRG offsets, the position and
velocity centroids of the galaxy groups defined in Table \ref{KMM3T} are
used to calculate the bound/unbound criteria.  As will be shown in section
\ref{sub:x-ray_data}, the group centroids correspond well with the peaks
seen in the X-ray data.

\placefigure{BNDAB}

\placefigure{BNDABC}

From Figures \ref{BNDAB} and \ref{BNDABC} a bound condition for groups 1 and 2
is satisfied between the projection angles of $20^{\circ}$ and $84^{\circ}$.
Similarly, group 3 is bound for projection angles between $24^{\circ}$ and
$81^{\circ}$.

\subsection{Dominant Galaxies}
\label{sub:dominant_galaxies}

Computer simulations by \citet{BODE94} indicate that a first ranked galaxy
settles into the center of a poor cluster of galaxies in a time less than
the Hubble time.  We shall define the FRG as the brightest galaxy within a
defined group.  The simulations show that the time that it takes this FRG to
settle into the center is dependent on the velocity dispersion of the
galaxies within the cluster.  As the velocity dispersion increases, the rate
of successful mergers decreases.  Bode \ea\ (1994) argue that their models
support the hypothesis that the FRG first forms in the central region of a
poor cluster.  In this environment, velocity dispersions are low and
successful mergers are frequent, allowing the efficient development of a
central dominant galaxy.

In Table \ref{BRIGHT} are listed the five brightest galaxies associated with
A2256.  The four brightest galaxies are found within the central region of
the cluster.  The fifth brightest galaxy is found centrally located within
subclump 3.  Figure \ref{DOMINANT} shows the velocity histogram with the
five brightest galaxies marked to indicate their position within the three
subclumps.  Not only are the respective galaxies closely associated with the
corresponding groups in positional space, but also in velocity space.  The
first and third galaxies are centrally located within the velocity
distributions of subclumps 1 and 2, respectively.  The fifth brightest is at
the center of the velocity histogram feature associated with clump 3.

\placetable{BRIGHT}

The second (ID \#530) and fourth (ID \#420) brightest galaxies are located
within the central region of the cluster.  Both galaxies are assigned to group
2 and have very similar velocities (see Table \ref{BRIGHT}).  These galaxies
occupy an intermediate position between groups 1 and 2 in velocity space.  One
can speculate as to the true origin of these galaxies.  Our merger hypothesis
suggests that they are either first-ranked galaxies from other poor groups
that merged at some time in the past, or that they are the second-ranked
galaxies associated with groups 1 and 2 that have not yet merged with the
central dominant galaxy of their respective group.  Once these galaxies enter
the environment of the merged system the velocity dispersion increases.  The
chance of a merger occurring between the the central dominant galaxies
decreases significantly, and the second and fourth brightest galaxies may not
merge with the first and third brightest galaxies for some time.  We do not
see evidence of groups corresponding to the second and fourth brightest
galaxies, so we favor the second interpretation.

\placefigure{DOMINANT}

\subsection{X-Ray Data}
\label{sub:x-ray_data}

Several X-ray maps of A2256 are available in the literature
\citep[e.g.][]{BRIEL91,BRIEL94,MARKEVITCH96}.  It is one of the best examples
of a cluster showing X-ray substructure.  The X-ray image shows two central
peaks, separated by about $3\arcmin\!\!.5$, that are located near the optical
center of the cluster \citep{BRIEL91}.  As has been widely discussed, the
X-ray emission can be used to trace the total gravitational potential of the
cluster, under the assumption that the X-ray emitting gas is in hydrostatic
equilibrium.

If the luminous matter of the galaxies also traces the gravitational
potential, then the substructures found by application of the EM algorithm
should approximately correspond to the structure seen in the X-ray surface
brightness distribution.  The positions of the centers of groups 1 and 2,
listed in Table \ref{KMM3T}, correspond well with the two maxima observed in
the extended cluster emission.  Briel \ea\ (1991) mark the center of the
central peak approximately $70 \arcsec$ to the west of the central cD galaxy
(ID \#581).  The center of group 2 is also approximately $70 \arcsec$ to the
northeast.  This places the center of group 2 approximately $10 \arcsec$ to
the northeast of the central cD galaxy.  The proximity of these positions
strongly suggests that group 2 is the dominant group.

The position of group 1 is coincident with the position of the excess source
defined by Briel \ea\ (1991).  According to Table \ref{BRIGHT}, the FRG for
group 1 (ID \#428) is located at RA: $17^{h}\,03^{m}\,35\fs6$, Dec:
$78^{\circ}\,37\arcmin\,45\farcs1$, placing it only about $2\farcm3$ southeast
of the excess source peak.  The center of the group 1 as obtained from the KMM
algorithm is only approximately $5 \arcmin$ to the west of this peak.  This
also suggests a correlation between the excess source and the smaller
infalling group 1.

\placefigure{OTHER}

The left-hand panels in Figure \ref{OTHER} show the position of each group
along with the X-ray contours \citep{BRIEL94}.  The central dominant galaxy
that corresponds to each group is marked by a star, and the centroids for
groups 1, 2 and 3 are marked by a ``X.''  However, the small number of
galaxies associated with group 3 make the position for the centroid for group
3 uncertain.  One should note that the centroid for group 3 is only $3.5
\arcmin$ to the northeast of the FRG for group 3 (ID \#298).

There are no other features in the X-ray surface brightness distribution that
indicate the presence of a third group.  This is not unexpected, since group 3
has a mass comparable to that of a poor group.  Thus, the X-ray luminosity of
group 3 is probably considerably less than those of group 1 and 2, and so the
X-ray emission of A2256 is dominated by these brighter groups.  Nevertheless,
in section \ref{sub:radio_data} we will see that there is potential evidence
in the radio emission for group 3.

\subsection{Radio Data}
\label{sub:radio_data}

In the right-hand panels of Figure \ref{OTHER}, we present radio contours
\citep{ROTTGERING94} of the central region of A2256, along with the positions
of the galaxies with measured redshifts.  We shall use the same naming
convention for features in the radio map as was used by \citet{ROTTGERING94}
and \citet{BRIDLE79}.  For convenience, Table \ref{RADIO} includes a subset of
the table provided by \citet{ROTTGERING94}.  Column (1) gives the feature
identification \citet{ROTTGERING94}, and Bridle \ea\ (1979).  Columns (2) and
(3) give the RA and Dec, and Column (4) give the optical counterpart if one is
known.  The identification number used in Column (4) corresponds with the
galaxy ID numbers used in Table \ref{CATALOG}.  Column (5) gives the group
membership of the optical counterpart if one is known, and column (6) gives
the measured radial velocity of the galaxy.

\placetable{RADIO}

There are a number of striking features in the radio map of A2256 that can be
easily separated into the effects from the interaction between groups 1 and 2
versus the interaction between groups 2 and 3.  We shall discuss these two
feature sets in turn.  To facilitate the discussion, Figure \ref{RADIO.PEAK}
shows the positions of the radio peaks superimposed on the radio contours from
\citet{ROTTGERING94}.  The positions adopted are those given in
\citet{ROTTGERING94}.

\placefigure{RADIO.PEAK}

\subsubsection{Head-Tail Radio Sources A \& B}
\label{subsub:radio_a_b}

We interpret the combined radio feature formed from peaks A and B as the
product of the merger event between the groups 1 and 2.  Table \ref{RADIO}
indicates that galaxies associated with group 2 are the optical counterparts
to peak A, and galaxies associated with group 1 are the counterparts to peak
B\@.  In our interpretation, group 1 is infalling radially from behind onto
group 2, from the northwest, with a position angle consistent with the value
given by \citet{ROETTIGER93}.  The radio feature A is most likely associated
with the high speed interaction between galaxies 410 and 420 of group 2.
Radio peak B is indisputably associated with galaxy 335 of group 1.  This
feature is one of four known head-tail sources.  In most models a head-tail
source is the trail of a radio-active galaxy that moves at high speeds
relative to the intracluster medium.  According to our model, source B is the
interaction of galaxy 335 with the intracluster medium of group 2 as the two
clusters merge.

\subsubsection{Diffuse Radio Emission Sources G \& H}
\label{subsub:radio_g_h}

The diffuse emission feature marked by peaks G and H was originally
interpreted as a shock feature between the merging groups 1 and 2 by
\citet{ENSSLIN98}.  However, we feel this feature provides strong evidence
that group 3 is bound and infalling on group 2.

In support of this interpretation, there is a strong correlation between the
location of the galaxies associated with group 3 as determined from the EM
algorithm and the extended radio feature (see Figure \ref{OTHER}).  The shape
and position angle of the galaxy distribution closely follow the contour lines
of the G and H radio feature.  Each galaxy associated with peaks directly
related to the extended radio feature is a member of either group 2 or 3 with
the central dominant galaxy associated with group 3 positioned central to the
G-H radio feature.  The only exception is peak J which is not directly related
to the G-H radio halo, but is associated with a cluster 1 member.

\citet{ROTTGERING94} show a strong correlation between the X-ray contours and
the northwest edge of the G-H radio feature.  Given the mass of group 3, the
expected X-ray thermal bremsstrahlung luminosity will be in the range of
$10^{41}$--$10^{42}$\ergss.  Furthermore, computer simulations of particles
accelerated by shocks formed between merging clusters of galaxies
\citep{BERRINGTON01} show that the expected X-ray luminosity from the
nonthermal particles accelerated by first-order Fermi in the $2$--$10$\kev\
energy range is $\sim\!\!10^{43}$\ergss at peak luminosity.  It is challenging
to distinguish between nonthermal X-ray emission from the shock front and
thermal bremsstrahlung emission from group 3 given the noise level of the
ROSAT imaging.  Emission from the shock front should dominate and more readily
manifest itself in the ROSAT data as supported by the analysis performed by
\citet{ROTTGERING94}.  More recent observations by the {\it Chandra} X-ray
Observatory \citep{SUN01} indicate an X-ray excess in the region of group 3
and the G-H radio feature.

\subsubsection{Head-Tail Radio Sources C \& I}
\label{subsub:radio_c_i}

Source C is unresolved laterally, but is approximately 500 kpc in length with
a kink in the tail 380 kpc distant from the source \citep{ROTTGERING94}.
Galaxy 414, the radio source C optical counterpart, is a member of group 2.
Our interpretation is the head-tail radio feature C has just passed through
the shock front marked by the G and H radio feature.  The kink marks the point
where the tail pops through the shock front into the intergalactic medium of
group 3.  The velocity of galaxy 414 deviates from the velocity centroid by
only $\sim\!\!108\ \kmsec$.  Therefore, the orbital motion of galaxy 414 with
respect to group 2 is mostly tangential.  \citet{ROTTGERING94} performed an
orbit calculation assuming the tail traces the trajectory of the galaxy
through the cluster.  They find the shape of the tail to be consistent with a
galaxy with initial transverse velocity in the range of 2000 to 3500 \kmsec.

The head-tail source I is associated with galaxy 88 in group 2.  This optical
counterpart is found $\sim\!\!340\ \kmsec$ blueward of the group 2 centroid.
There are no galaxies associated with any obvious interaction with galaxy 88.
One possible interpretation is that it is a galaxy that has recently passed
from the intergalactic medium of group 2 and entered the intergalactic medium
of group 3.  Supporting this idea, the tail of source I extends only to the
diffuse emission nebula source of G and H.  The radio sources C and I along
with sources G and H provide strong evidence for interaction between groups 2
\& 3 (see \S \ref{sec:cluster_dynamics}).

\section{Summary and Conclusion}
\label{sec:summary_and_conclusion}

We have presented the results of a radial velocity survey of the galaxy
cluster A2256, carried out with the Hydra multi-object spectrograph on the
WIYN telescope, in which we obtained a total of 277 radial velocities of
cluster members.  In addition, we presented new photometry and astrometry of
861 cluster members within the central $19\arcmin \times 19\arcmin$ of the
cluster, obtained from a $3 \times 3$ WIYN image mosaic in the R band.  By
applying a number of statistical tests, we found strong evidence of
statistically significant deviations from the expected single-Gaussian
velocity distribution of a relaxed cluster.  On application of the KMM
software package, which is based on the EM algorithm for mixture modeling
\citep{MCLACHLAN88}, we found a total of three subclusters, only two of which
had been previously noted.  This analysis used both velocity and spatial
distribution data to identify the separate groups.  We obtained virial mass
estimates for each group and showed that the total of the group masses is less
than half the inferred virial mass of the entire cluster, in support of the
presence of substructure.  We then applied the two-body orbit model of Beers
\ea\ (1982), first to the two-body system defined by groups 1 and 2 alone and
then to a system in which groups 1 and 2 were treated as a single body and
group 3 was taken as the other body.  This dynamical analysis strongly
indicates that groups 1 and 2 are an infalling, bound system.  However it does
not, in itself, fix the dynamical status of group 3 within the A2256 system.

To help determine whether group 3 is bound to the main cluster, we correlated
our optical results with previous X-ray and radio imaging of A2256.  The X-ray
surface brightness distribution of A2256 has two centrally located peaks
\citep{BRIEL91}.  The galaxy groups 1 and 2, which we identified from the
galaxy position and velocity distributions, are strongly correlated with the
X-ray peaks, providing additional support for the merging subcluster
interpretation.  The optical data alone allow group 3 to be modeled as either
bound, and thus in close proximity with the 1--2 system, or unbound, and thus
not interacting with 1--2 system.  However, radio maps of A2256
\citep{ROTTGERING94} provide strong evidence for an interaction between groups
2 and 3, and thus clearly support the bound model.  As discussed in section
\ref{sub:radio_data}, the diffuse radio emission feature is probably a shock
front between the intracluster media of groups 2 and 3.  In addition, the
interaction of the head-tail sources C and I with the diffuse radio emission
feature provides evidence that the two systems are interacting (see
\S\ref{subsub:radio_g_h}).

In summary, our proposed model of A2256 consists of three bound subgroups that
are undergoing merging.  Groups 1 and 2 appear to be in an advanced stage of
the merging process, near the time of the first close passage of the group
centers.  In this model, group 1 is located behind group 2 and is infalling
from the northwest.  Pinkney \ea\ (1996) suggest that the line of centers lies
at approximately 45\arcdeg\ from the line of sight, which is consistent with
the range allowed for bound orbits.  According to our model, group 3 is
located on the near side of the 1--2 system and is infalling from the north.
Our two-body analysis indicates that the infall direction lies at an angle to
the line of sight in the range 50\arcdeg\ -- 60\arcdeg.  The dynamics of the
dominant group 2 are probably strongly influenced by the ongoing merger with
group 1, but the 1--2 merger system may be largely unaffected by the presence
of the much smaller group 3.  Each group contains a central, dominant galaxy
which, for groups 1 and 2, is located near the X-ray surface brightness peak.
This is consistent with the prediction that the dominant galaxy of a group
will quickly settle into the center of the local potential well
\citep{BODE94}.  The greatest offset between the galaxy distribution centroid,
the position of the central dominant galaxy, and the X-ray peak occurs in
group 1, suggesting that the apparent merger event is strongly influencing the
dynamics of this group.

As discussed by \citet{HENRIKSEN99}, there is some continuing controversy
regarding the merger model for interpreting the X-ray observations of A2256.
Thus, further X-ray observations of A2256 would help to distinguish between
the cases of merging subclusters versus clusters that are merely superimposed
on the sky by projection \citep{HENRIKSEN99}.  Briel \& Henry (1994) suggest
the presence of a complex temperature structure in the X-ray intensity
distribution, but this is not strongly supported by observations made by ASCA
\citep{MARKEVITCH96}.  Because of the high velocity impact between groups 1
and 2, the presence of a shock front is likely and has been predicted by
computer simulations \citep{ROETTIGER93}.  A2256 is an excellent target for
the Chandra X-ray Observatory.  Higher resolution X-ray images and better
temperature maps of the A2256 field would greatly increase our understanding
of the dynamics of the multiple merger events that appear to be shaping this
system.

\acknowledgements

We thank Mike Pierce for many long and useful discussions concerning the data
acquisition, reduction and interpretation.  We thank Ulrich Briel and J.\
Patrick Henry for the use of the ROSAT X-ray data in Figure \ref{OTHER}.  We
thank Huub R\"{o}ttgering and Ignas Snellen for the use of the VLA radio data
in Figure \ref{OTHER}.  We also thank the Indiana University Astronomy
department for travel support to carry out the necessary observations at the
WIYN telescope.

\newpage

\figcaption[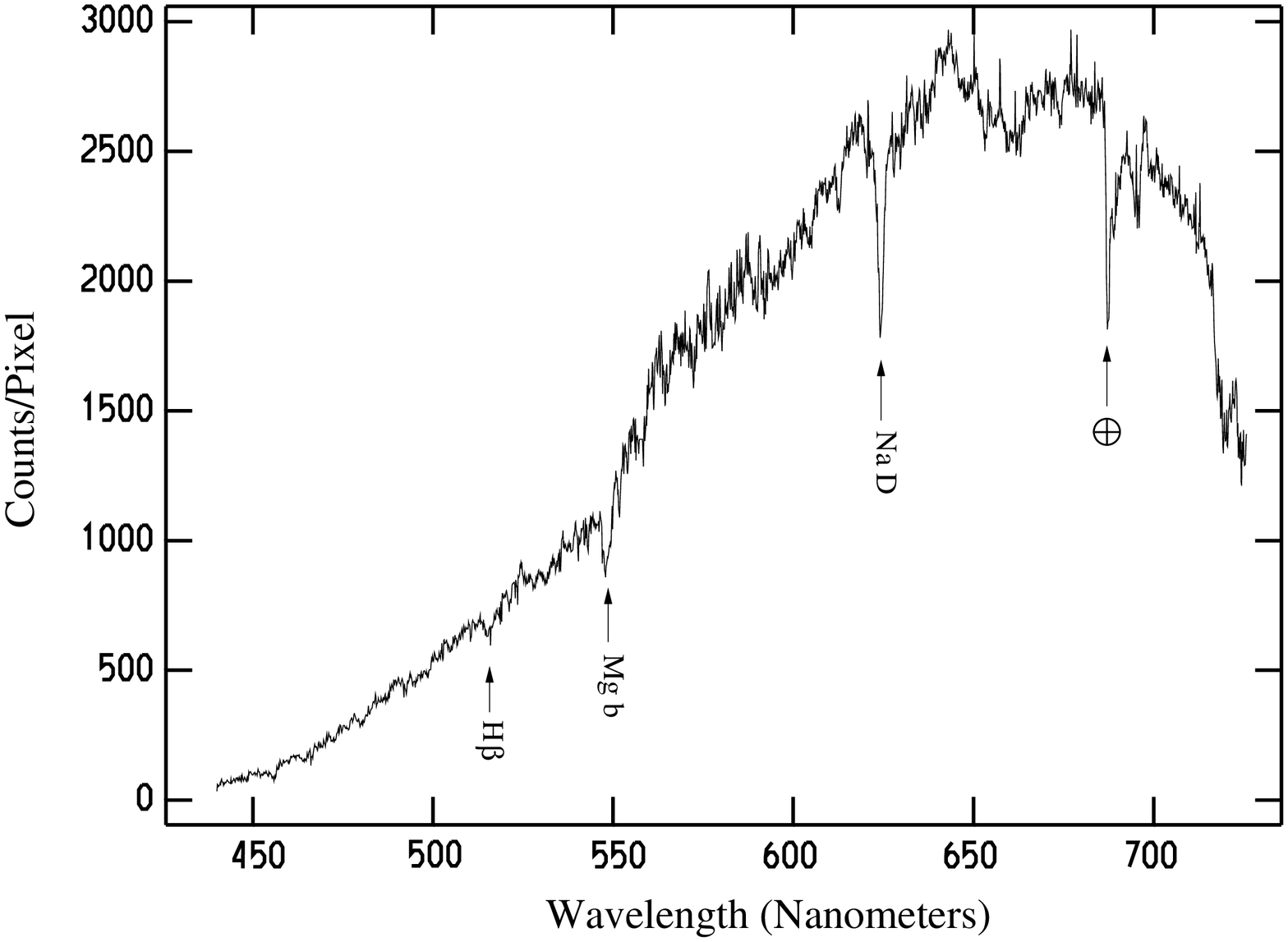]{Example reduced galaxy spectrum with the
target lines marked.  H$\alpha$ is often redshifted into the teluric feature
at 680 nm, as is the case here.\label{SPECTRA}}

\figcaption[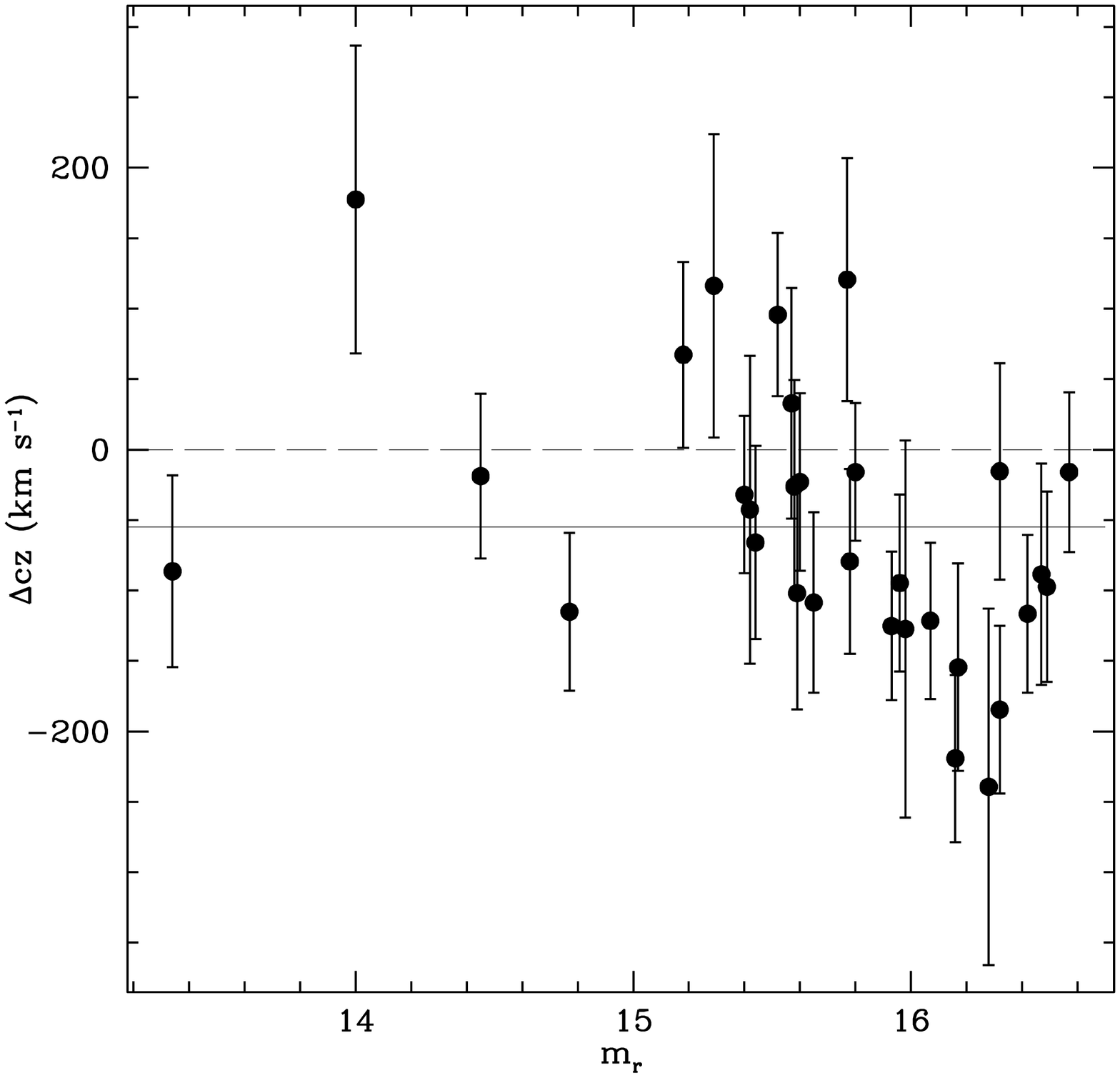]{Difference of Fabricant \ea\ (1989) and
our velocities from the June 1996 run.  We find a mean difference of 55
\kmsec\ with a standard deviation of 98 \kmsec.  The error bars are $1 \sigma$
error bars as calculated from the quadratic sum of the two
uncertainties.\label{DELTACZ}}

\figcaption[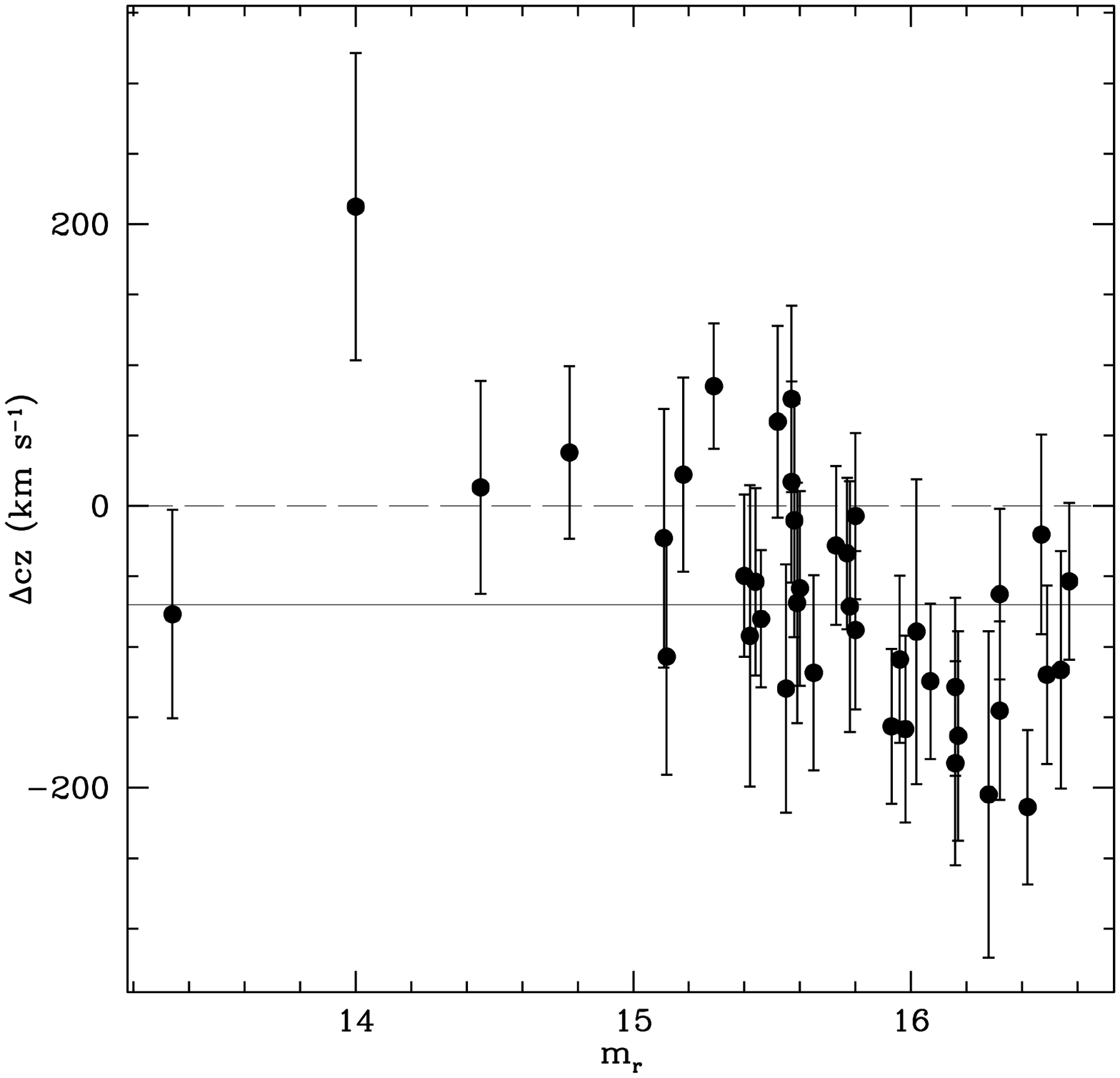]{Difference of Fabricant \ea\ (1989) and our
velocities from the June 1997 and June 1998 run.  We find a mean difference of
70 \kmsec\ with a standard deviation of 84 \kmsec.  The error bars are $1
\sigma$ error bars as calculated from the quadratic sum of the two
uncertainties.\label{BERRDELTACZ}}

\figcaption[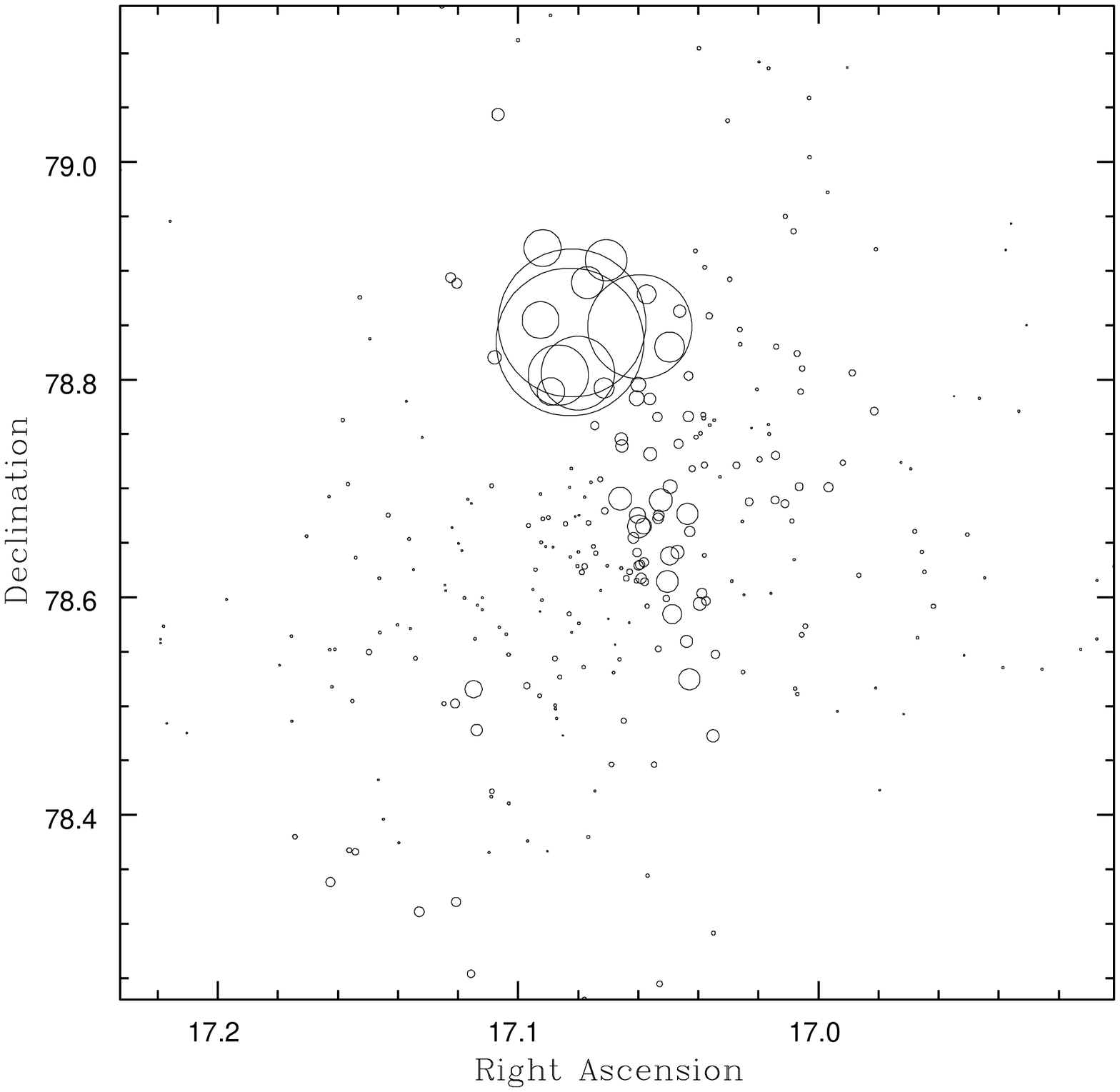]{Shows fraction of galaxies with measured
radial velocities as a function of R\KC\ magnitude.  Only galaxies with
measured magnitudes are included in the figure.  All magnitudes from
\citet{FAB89} were converted to R\KC\ magnitude.  As is shown above, the sample
included in table \ref{CATALOG} include about 70\% of the galaxies out to
$18^{th}$ magnitude.\label{COMPLETE}}

\figcaption[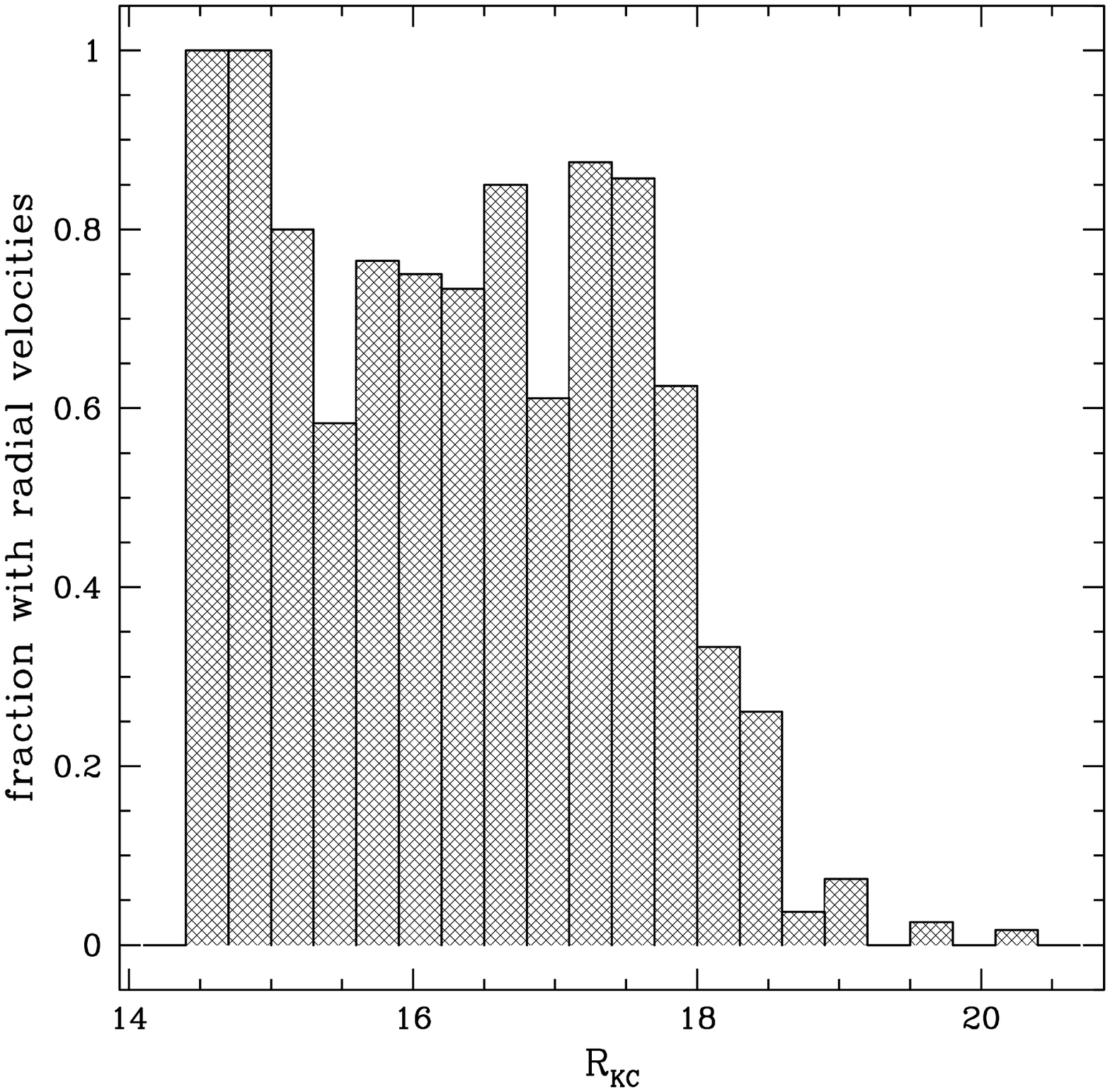]{Dressler--Shectman plot for A2256.  Each open
circle is a galaxy whose radius is proportional to the $e^{\delta}$.  $\delta$
is described by Dressler and Shectman as the deviation of the local mean and
sigma from the global mean and sigma.  See text for an
explanation.\label{DELTA}}

\figcaption[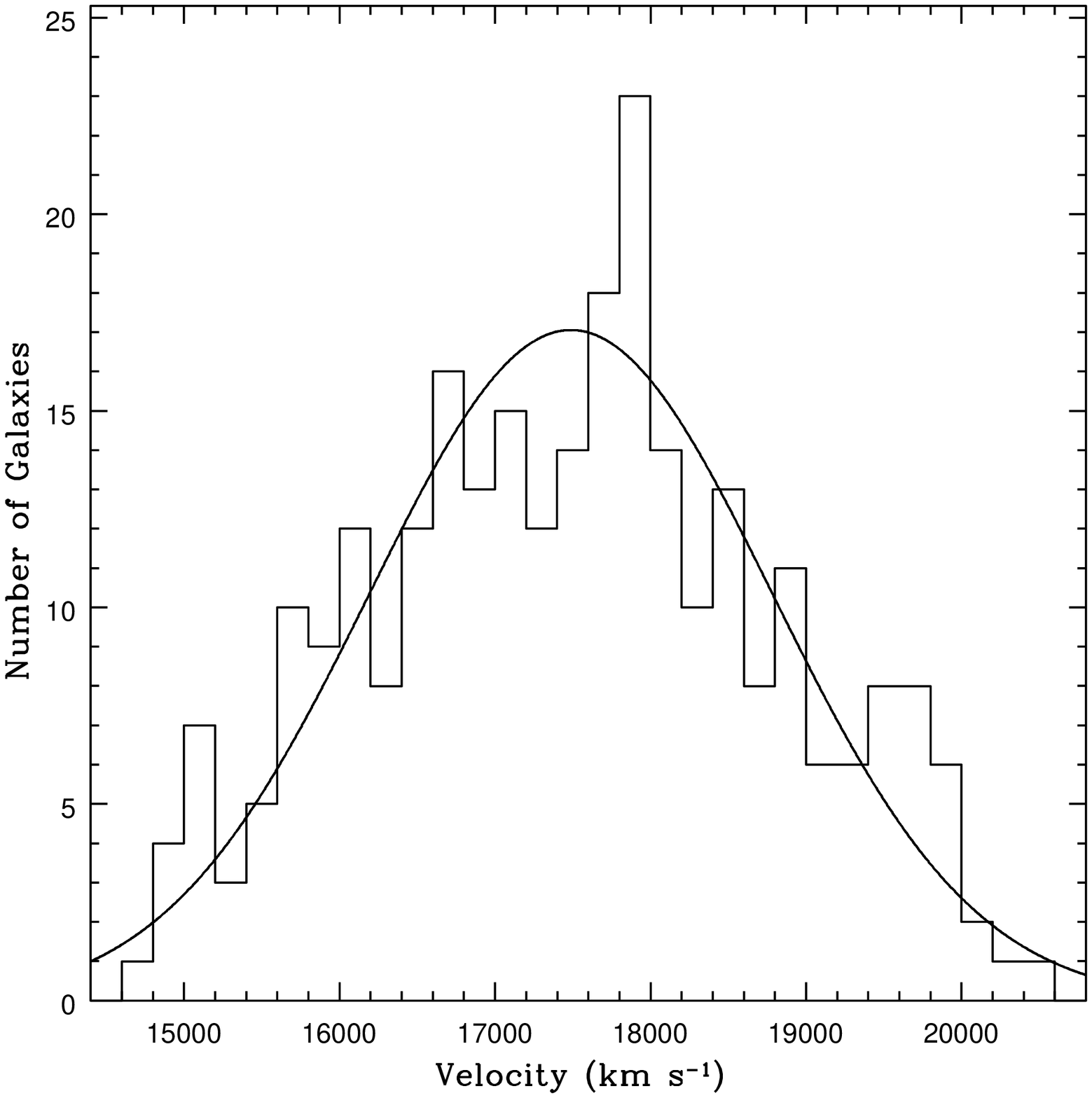]{Velocity histogram for A2256.  The solid line
is the best single-Gaussian fit to the data, with a mean of 17488.1 \kmsec\
and a standard deviation of 1295.9 \kmsec.\label{KMM1}}

\figcaption[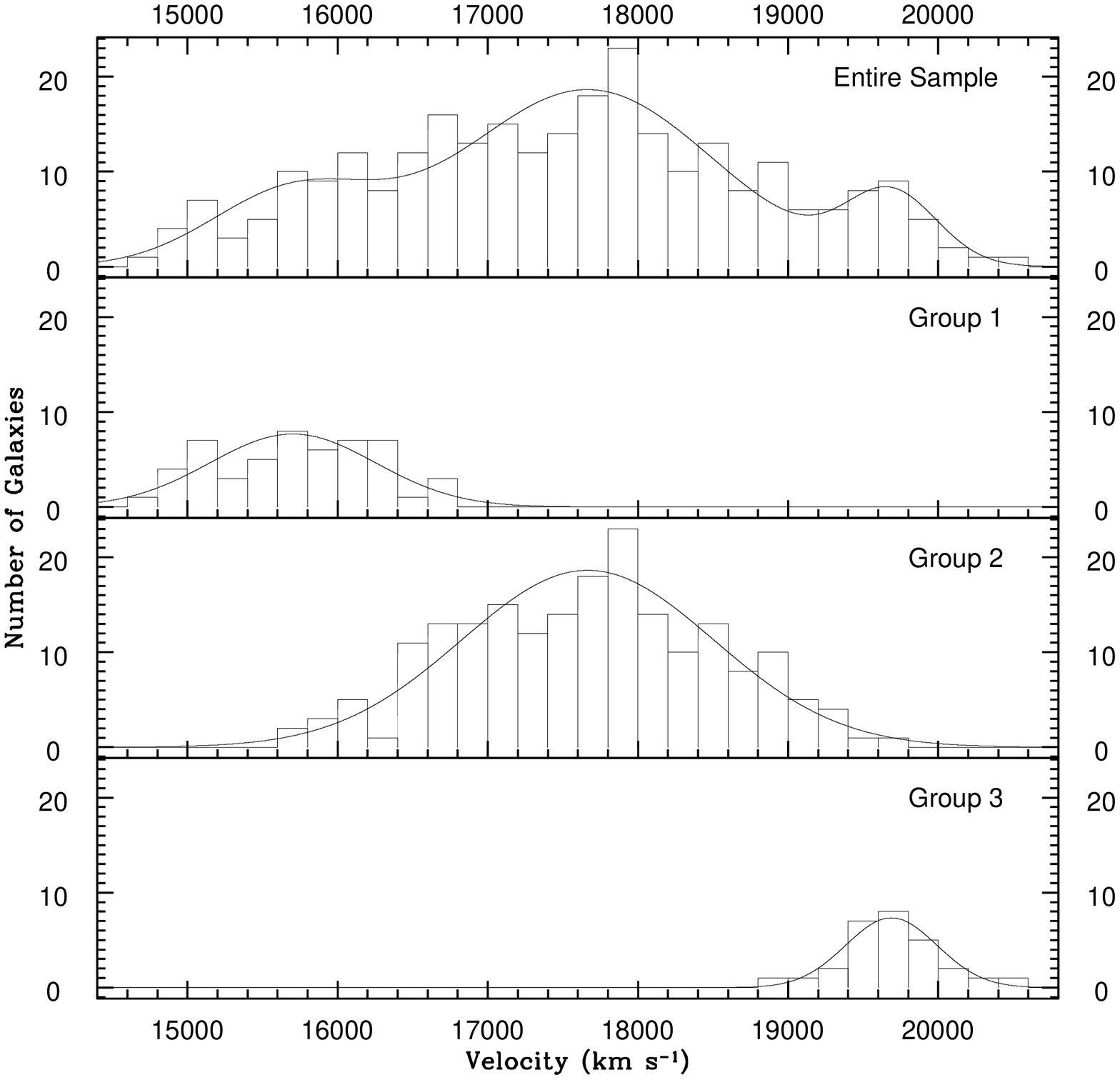]{Velocity histogram for A2256 with the
multiple Gaussian profiles.  The top panel shows the combined profile from the
multiple Gaussian model.  The second, third, and bottom panels show the
galaxies belonging to the appropriate groups 1, 2, and 3, respectively, along
with the best fit Gaussian.  Parameters for each Gaussian is described in
Table \ref{KMM3T}.\label{3GAU}}

\figcaption[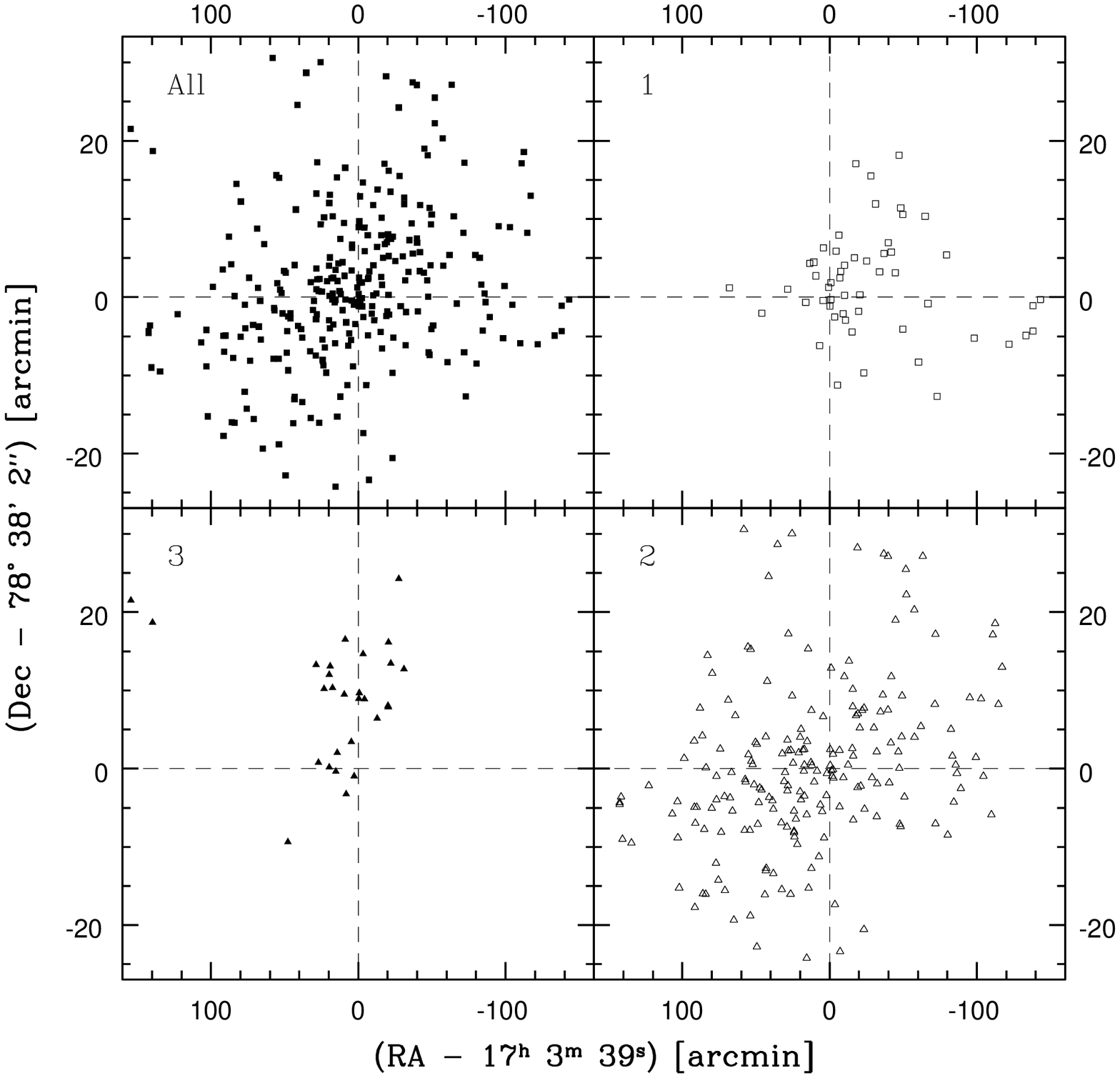]{Sky plot of the galaxies associated with the
three groups found by the KMM algorithm.  The open squares represent galaxies
that belong to group 1.  The open triangles represent galaxies that belong to
group 2.  The filled triangles are galaxies belonging to group 3.  The x and y
axes give the offsets in arcminutes from the central brightest galaxy (ID
\#581).\label{A2256radec4}}

\figcaption[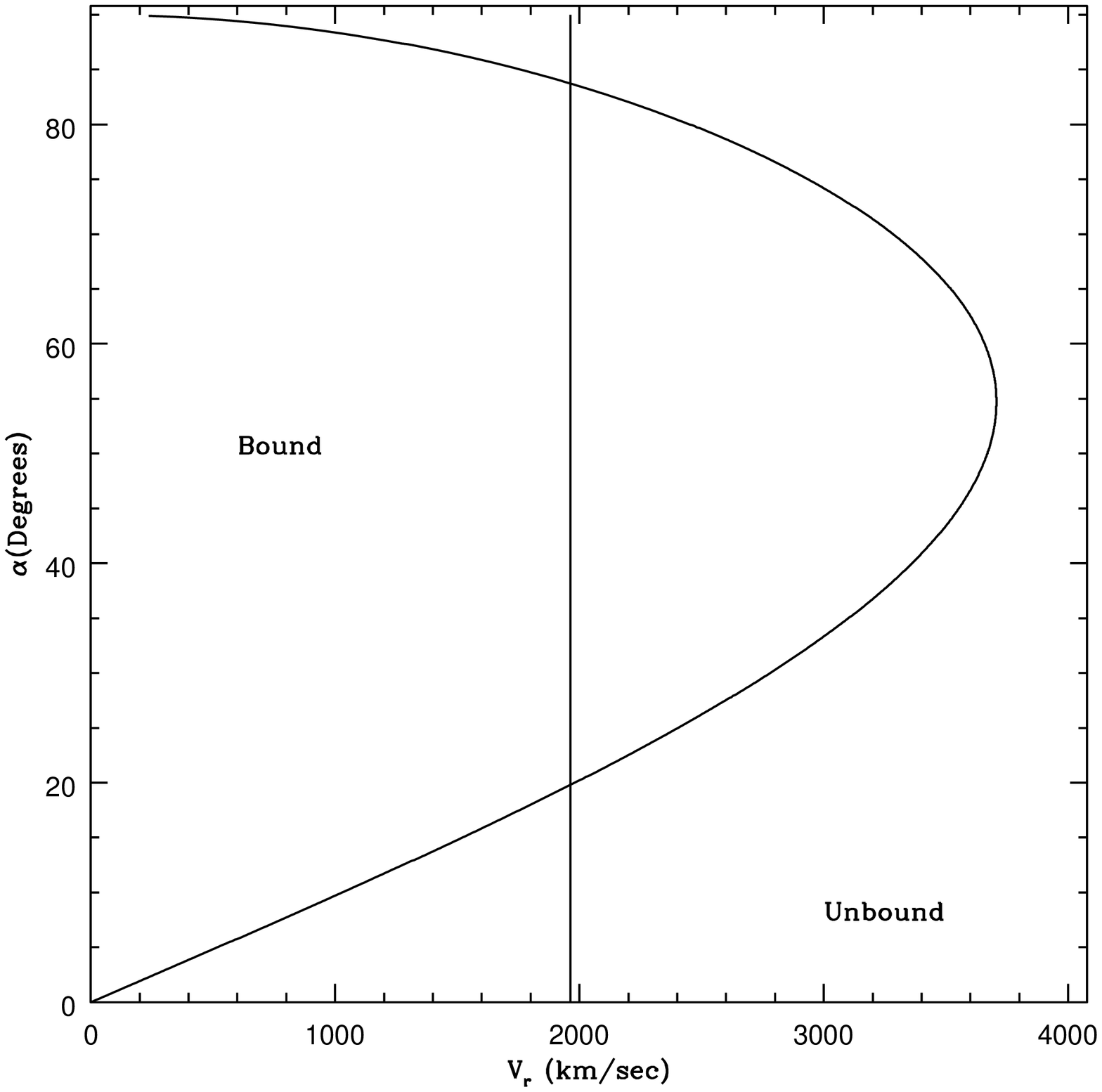]{Plot showing the boundary between the bound
and unbound orbits for Groups 1 and 2.  $V_{r}$ is the difference in the
radial velocities between the two subclumps (see text).  The angle between the
sky plane and the line joining the centers of clusters 1 and 2 is $\alpha$.
The vertical line marks the relative velocity difference between the two
groups.\label{BNDAB}}

\figcaption[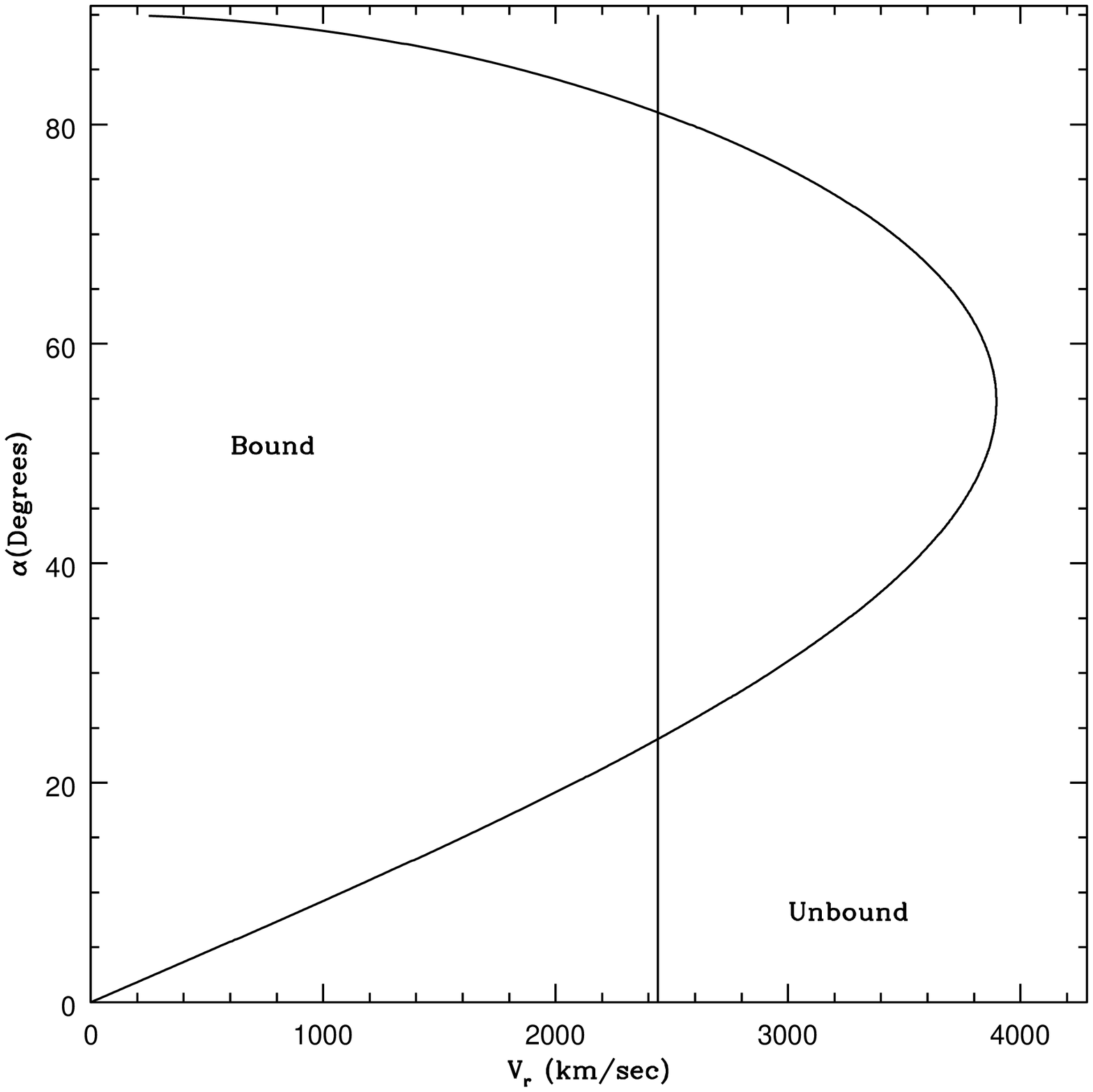]{Same as Figure \ref{BNDAB} with the exception
that the two groups are the combination of group 1 and 2 (group A), and group
3 (group B).\label{BNDABC}}

\figcaption[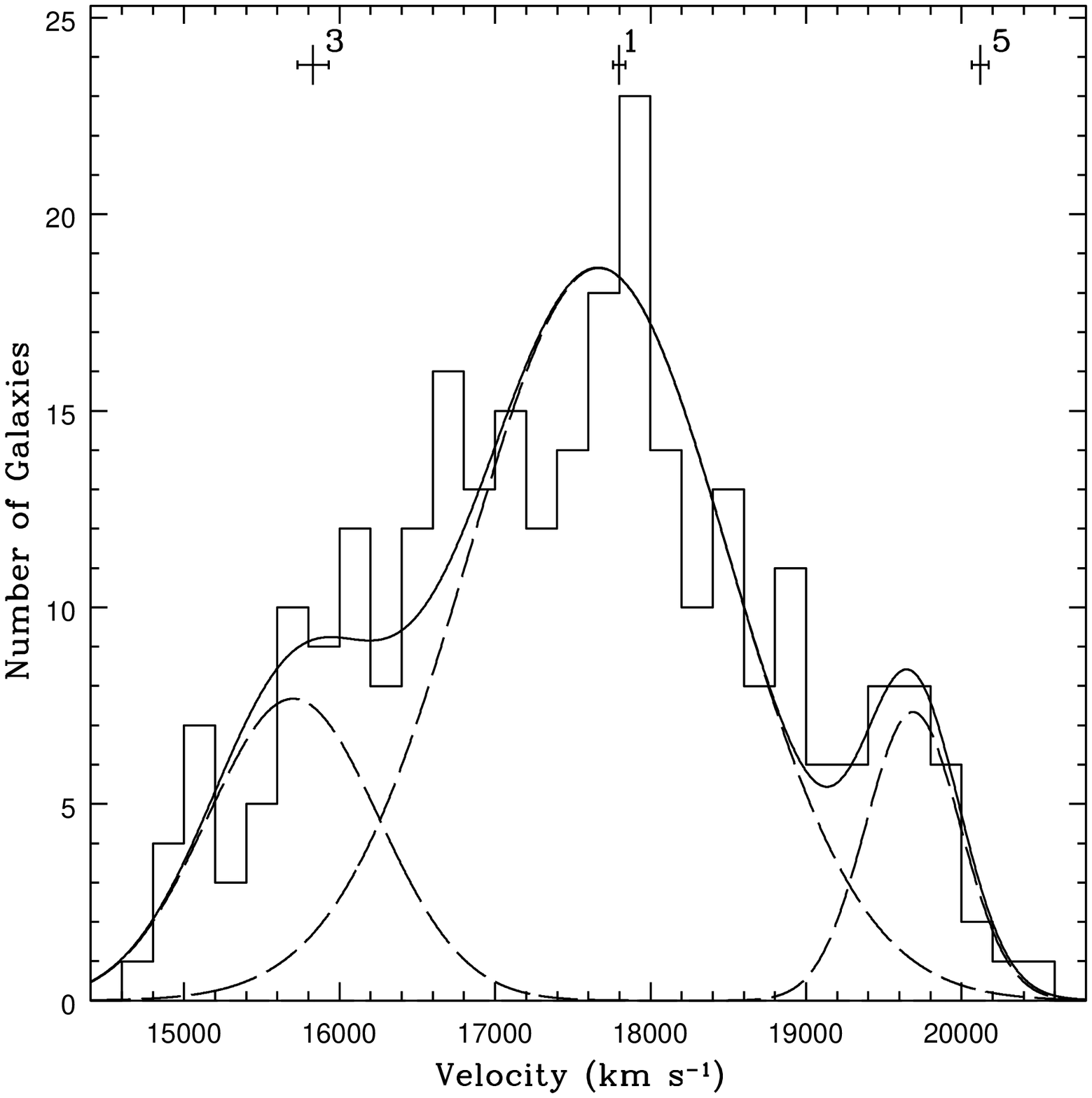]{Histogram of the galaxies that are a member
of A2256 with the 3-Gaussian model determined by application of the EM
algorithm.  The solid line shows the summed distribution of the combined
system.  The dashed line shows the individual Gaussians given in Table
\ref{KMM3T}.  The velocity locations of the first, third and fifth brightest
galaxies, from Table \ref{BRIGHT}, are indicated at the top of the diagram
with $1\sigma$ error bars.\label{DOMINANT}}

\figcaption[berrington.fig12.eps]{shows the positions of galaxies for groups
1, 2, and 3 along with the X-ray (left) and radio contours (right).  The top
row shows the galaxies associated with group 1; the second row shows galaxies
associated with group 2; and the bottom row shows the galaxies associated with
group 3.\label{OTHER}}

\figcaption[berrington.fig13.eps]{shows the positions of the radio peaks along
with the radio contour map from \citet{ROTTGERING94} in the central region of
A2256.  The same naming convention used in \citet{BRIDLE79} and
\citet{ROTTGERING94} for simplicity.  See Table \ref{RADIO} for celestial
coordinates and optical counterparts.  \label{RADIO.PEAK}}

\newpage

\pagestyle{empty}

%
%

\begin{deluxetable}{cccc}
\tablecolumns{4}
\tablewidth{0pt}
\tablecaption{Field centers\label{FIELDS}}
\tablehead{
	\colhead{Field}	&
	\colhead{RA}	&
	\colhead{Dec}	&
	\colhead{Run}	\\
	\colhead{}	&
	\multicolumn{2}{c}{(J2000)}	&
	\colhead{}	
}
\startdata
C3    & 17:06:07.5 & 78:44:25.0 & June 1996 \\
C4    & 17:06:08.3 & 78:38:25.0 & June 1996 \\
C5    & 17:05:58.0 & 78:32:50.7 & July 1997 \\
D3    & 17:03:52.0 & 78:44:11.0 & June 1996 \\
D4    & 17:04:01.8 & 78:38:11.4 & June 1996 \\
D5    & 17:04:04.2 & 78:31:41.7 & July 1997 \\
E3    & 17:01:41.0 & 78:44:00.0 & June 1996 \\
E4    & 17:01:45.4 & 78:37:29.0 & June 1996 \\
E5    & 17:01:40.0 & 78:32:40.0 & June 1996 \\
\enddata
\end{deluxetable}

\begin{deluxetable}{ccccc}
\tablewidth{0pt}
\tablecaption{Hydra Observation log\label{HYDRASETUP}}
\tablehead{
	\colhead{Observation Date}	&
	\colhead{Grating}	&
	\colhead{Central $\lambda$}	&
	\colhead{Fibers}	&
	\colhead{Exposure}	\\
	\colhead{}	&
	\colhead{}	&
	\colhead{(nm)}	&
	\colhead{}	&
	\colhead{(sec)}	
}
\startdata
June 1996 & 360@7.0  & 750.0 & Red  & $3\times1800$ \\
June 1997 & 600@13.9 & 580.0 & Blue & $2\times3600$ \\
June 1998 & 600@13.9 & 570.0 & Blue & $2\times3600$ \\
\enddata
\end{deluxetable}

\begin{deluxetable}{cccccccc}
\tablecolumns{8}
\tablewidth{0pt}
\tablecaption{Galaxy Catalog for A2256\label{CATALOG}}
\tablehead{
	\colhead{}	&
	\colhead{}	&
	\colhead{}	&
	\colhead{}	&
	\colhead{}	&
	\colhead{}	&
	\colhead{$1\sigma$}	&
	\colhead{}	\\
	\colhead{}	&
	\colhead{}	&
	\colhead{}	&
	\colhead{}	&
	\colhead{}	&
	\colhead{}	&
	\colhead{Velocity}	&
	\colhead{Fabricant}	\\
	\colhead{Number}	&
	\colhead{RA}	&
	\colhead{Dec}	&
	\colhead{$R\KC$}	&
	\colhead{$m_r$}	&
	\colhead{$\rm{cz}$}	&
	\colhead{Error}	&
	\colhead{Number}	\\
	\colhead{}	&
	\multicolumn{2}{c}{(J2000.0)}	&
	\colhead{}	&
	\colhead{}	&
	\multicolumn{2}{c}{(\kmsec)}	&
	\colhead{}	
}
\startdata
10  & 16:54:05.8 & 78:37:42.8 &                & 14.66 & 16710.4 & 36.4 & 11 \\
13  & 16:54:26.2 & 78:36:56.2 &                &       & 16649.1 & 47.9 &    \\
15  & 16:54:26.3 & 78:33:41.9 &                & 14.72 & 16197.4 & 44.7 & 14 \\
17  & 16:54:45.5 & 78:33:07.8 &                &       & 16032.4 & 47.3 &    \\
19  & 16:55:11.5 & 78:30:16.6 &                &       & 59267.5 & 49.2 &    \\
20  & 16:55:32.0 & 78:32:02.3 &                & 16.16 & 16179.6 & 34.9 & 17 \\
22  & 16:55:50.5 & 78:51:00.1 &                &       & 18346.6 & 44.8 &    \\
23  & 16:55:59.8 & 78:46:15.7 &                & 16.05 & 18436.9 & 52.0 & 19 \\
24  & 16:56:09.2 & 78:56:36.2 &                & 15.75 & 17851.4 & 46.3 & 20 \\
25  & 16:56:15.5 & 78:55:09.1 &                &       & 18308.9 & 47.7 &    \\
26  & 16:56:18.8 & 78:32:07.2 &                &       & 18374.3 & 33.7 &    \\
28  & 16:56:40.9 & 78:37:04.5 &                &       & 17867.8 & 33.4 &    \\
29  & 16:56:47.2 & 78:46:58.5 &                & 16.40 & 16947.3 & 46.3 & 23 \\
30  & 16:56:48.2 & 78:31:30.0 &                &       & 59465.4 & 58.1 &    \\
31  & 16:56:56.6 & 78:58:57.5 &                &       & 57833.1 & 52.5 &    \\
33  & 16:57:01.6 & 78:39:27.6 &                & 16.58 & 18835.7 & 46.8 & 24 \\
34  & 16:57:05.5 & 78:32:48.1 &                & 16.04 & 16494.7 & 31.0 & 25 \\
35  & 16:57:14.8 & 78:21:20.4 &                &       & 59241.4 & 59.3 &    \\
36  & 16:57:17.5 & 78:47:05.1 &                &       & 18199.9 & 34.9 &    \\
37  & 16:57:42.2 & 78:35:30.7 &                &       & 19386.7 & 40.9 &    \\
38  & 16:57:53.1 & 78:37:24.3 &                &       & 17432.1 & 47.6 &    \\
39  & 16:57:56.1 & 78:38:30.1 &                &       & 18085.7 & 57.6 &    \\
40  & 16:57:58.7 & 78:36:14.9 &                & 15.77 &  4253.1 & 44.6 & 27 \\
41  & 16:58:01.2 & 78:33:46.0 &                & 15.40 & 19202.0 & 40.0 & 28 \\
42  & 16:58:04.7 & 78:39:38.2 &                &       & 17975.8 & 39.4 &    \\
43  & 16:58:09.3 & 78:43:05.0 &                &       & 17143.6 & 46.1 &    \\
45  & 16:58:18.0 & 78:29:33.8 &                & 14.77 & 17881.9 & 46.5 & 29 \\
46  & 16:58:20.8 & 78:43:26.1 &                &       & 15909.3 & 51.6 &    \\
47  & 16:58:42.4 & 79:07:19.3 &                &       & 53343.8 & 60.9 &    \\
48  & 16:58:46.6 & 78:25:22.0 &                & 15.18 & 15635.2 & 49.2 & 30 \\
49  & 16:58:51.1 & 78:55:11.6 &                &       & 18086.7 & 47.6 &    \\
50  & 16:58:51.4 & 78:30:59.5 &                & 16.11 & 18504.0 & 44.0 & 31 \\
51  & 16:58:53.0 & 78:46:16.1 &                &       & 19347.7 & 66.1 &    \\
52  & 16:59:11.5 & 78:37:13.1 &                &       & 16629.6 & 42.9 &    \\
53  & 16:59:12.6 & 78:33:32.7 &                &       & 59325.7 & 68.0 &    \\
54  & 16:59:19.4 & 78:48:22.6 &                &       & 15126.3 & 38.5 &    \\
55  & 16:59:25.6 & 79:05:12.6 &                & 16.22 & 18143.1 & 57.1 & 33 \\
56  & 16:59:30.8 & 78:43:25.1 &                &       & 16943.7 & 37.3 &    \\
57  & 16:59:37.2 & 78:29:43.1 &                & 16.17 & 15639.8 & 55.0 & 32 \\
58  & 16:59:42.8 & 78:37:08.6 &                &       & 53413.1 & 49.8 &    \\
59  & 16:59:48.0 & 78:42:04.3 &                &       & 16595.6 & 48.9 &    \\
60  & 16:59:49.0 & 78:58:19.8 &                &       & 17638.5 & 36.4 &    \\
61  & 17:00:09.7 & 78:59:25.7 &                &       & 52218.3 & 108.0&    \\
62  & 17:00:10.7 & 79:00:15.8 &                & 16.44 & 17414.7 & 69.7 & 37 \\
63  & 17:00:11.2 & 79:03:32.1 &                &       & 17667.5 & 34.2 &    \\
64  & 17:00:15.7 & 78:34:24.5 &                & 16.30 & 18355.0 & 62.0 & 34 \\
65  & 17:00:19.5 & 78:48:37.1 &                & 16.13 & 15332.0 & 47.0 & 38 \\
66  & 17:00:20.0 & 78:33:55.7 &                & 16.71 & 16372.4 & 55.4 & 35 \\
67  & 17:00:21.3 & 78:47:20.2 &                & 16.12 & 18931.0 & 32.0 & 40 \\
68  & 17:00:23.1 & 78:42:06.2 &                & 16.77 & 17637.9 & 29.4 & 39 \\
69  & 17:00:25.4 & 78:30:39.8 &                & 16.32 & 17358.3 & 30.8 & 36 \\
70  & 17:00:25.5 & 78:49:26.6 &                &       & 15375.6 & 51.7 &    \\
71  & 17:00:27.9 & 78:30:57.9 &                &       & 18153.8 & 29.7 &    \\
72  & 17:00:29.2 & 78:38:04.6 &                & 16.40 & 17775.0 & 54.0 & 41 \\
73  & 17:00:29.9 & 78:56:10.9 &                & 16.39 & 15237.0 & 47.0 & 43 \\
74  & 17:00:31.7 & 78:40:12.3 &                & 15.77 & 17887.3 & 40.7 & 42 \\
79  & 17:00:39.7 & 78:57:00.1 &                &       & 18383.2 & 44.7 &    \\
80  & 17:00:40.2 & 78:41:10.0 & 16.77$\pm$0.08 &       & 16284.2 & 33.8 &    \\
86  & 17:00:50.7 & 78:49:49.3 &             & 16.31 & 17357.0 & 43.0 & 45 \\
87  & 17:00:51.3 & 78:43:49.2 & 18.10$\pm$0.08 &       & 15916.8 & 42.3 &    \\
88  & 17:00:52.0 & 78:41:21.9 & 15.39$\pm$0.07 & 15.89 & 17325.0 & 39.0 & 44 \\
96  & 17:00:56.9 & 78:36:13.0 & 16.74$\pm$0.07 &       & 18281.5 & 32.4 &    \\
99  & 17:00:59.0 & 78:45:00.1 & 15.29$\pm$0.08 & 15.58 & 16136.7 & 52.8 & 46 \\
101 & 17:00:59.7 & 79:05:10.0 &                & 16.58 & 18233.8 & 47.7 & 48 \\
102 & 17:00:59.9 & 78:45:31.8 & 17.31$\pm$0.08 &       & 17027.8 & 46.8 &    \\
116 & 17:01:10.5 & 78:43:36.3 & 15.08$\pm$0.08 & 15.29 & 16122.9 & 43.3 & 47 \\
119 & 17:01:11.2 & 79:05:31.0 &                & 16.18 & 18538.2 & 31.4 & 50 \\
125 & 17:01:13.8 & 78:47:27.9 &                &       & 18590.7 & 83.6 &    \\
138 & 17:01:20.2 & 78:45:19.8 & 18.02$\pm$0.08 &       & 16969.8 & 15.1 &    \\
142 & 17:01:23.1 & 78:41:15.8 & 16.28$\pm$0.08 & 15.73 & 16217.0 & 67.0 & 49 \\
152 & 17:01:29.4 & 78:36:08.0 & 15.77$\pm$0.07 & 16.16 & 17963.5 & 42.3 & 52 \\
154 & 17:01:30.4 & 78:31:52.5 & 16.73$\pm$0.08 & 16.69 & 17674.6 & 67.6 & 51 \\
155 & 17:01:31.2 & 78:40:10.9 & 19.38$\pm$0.07 &       & 19673.6 & 15.3 &    \\
157 & 17:01:33.8 & 78:49:57.3 &                &       & 15668.7 & 43.3 &    \\
158 & 17:01:34.3 & 78:50:45.9 &                &       & 19518.4 & 15.1 &    \\
166 & 17:01:38.2 & 78:43:16.7 & 17.09$\pm$0.08 &       & 17065.0 & 83.2 &    \\
178 & 17:01:43.8 & 78:36:53.6 & 17.08$\pm$0.07 &       & 18096.1 & 85.0 &    \\
179 & 17:01:44.4 & 78:46:06.2 & 17.86$\pm$0.08 &       & 51759.5 & 93.7 &    \\
183 & 17:01:46.4 & 78:53:32.0 &                &       & 14864.3 & 36.5 &    \\
186 & 17:01:47.5 & 78:51:23.9 &                &       & 34075.9 & 44.8 &    \\
187 & 17:01:48.8 & 79:02:16.8 &                & 13.79 & 19302.7 & 59.6 & 54 \\
200 & 17:01:58.1 & 78:42:38.4 & 16.65$\pm$0.08 &       & 14834.6 & 40.5 &    \\
205 & 17:02:03.5 & 78:32:51.4 & 17.87$\pm$0.08 &       & 17506.9 & 37.2 &    \\
207 & 17:02:05.0 & 78:45:46.0 & 15.53$\pm$0.08 & 15.78 & 17698.6 & 76.2 & 55 \\
208 & 17:02:05.8 & 78:17:29.1 &                & 15.55 & 18001.0 & 47.0 & 53 \\
209 & 17:02:06.3 & 78:28:21.6 &                &       & 16249.2 & 82.3 &    \\
215 & 17:02:10.3 & 78:45:29.6 & 15.78$\pm$0.08 & 16.02 & 17988.0 & 43.0 & 57 \\
217 & 17:02:10.9 & 78:51:31.0 &                & 16.12 & 19421.0 & 59.0 & 59 \\
225 & 17:02:14.7 & 78:35:47.7 & 15.87$\pm$0.08 & 15.64 & 18756.9 & 39.3 & 56 \\
228 & 17:02:16.3 & 78:54:11.6 &                &       & 19559.3 & 48.2 &    \\
230 & 17:02:16.7 & 78:43:17.4 & 16.05$\pm$0.08 & 16.56 & 17991.6 & 37.7 & 60 \\
231 & 17:02:16.8 & 78:38:18.7 & 16.22$\pm$0.07 & 16.67 & 15678.9 & 41.3 & 58 \\
232 & 17:02:17.5 & 78:45:52.6 & 16.29$\pm$0.08 & 16.74 & 19907.4 & 66.3 & 61 \\
233 & 17:02:18.1 & 78:46:03.8 & 15.62$\pm$0.08 & 15.96 & 19643.0 & 43.8 & 63 \\
236 & 17:02:19.8 & 78:36:13.1 & 16.08$\pm$0.07 &       & 15577.5 & 75.7 &    \\
240 & 17:02:21.3 & 78:45:02.4 & 16.82$\pm$0.08 &       & 17550.1 & 33.5 &    \\
242 & 17:02:22.4 & 78:35:38.5 & 16.16$\pm$0.08 & 16.61 & 16705.0 & 34.2 & 62 \\
243 & 17:02:23.3 & 79:06:16.3 &                &       & 17300.0 & 46.0 &    \\
249 & 17:02:26.6 & 78:44:49.9 & 16.56$\pm$0.08 &       & 16865.1 & 36.4 &    \\
251 & 17:02:27.5 & 78:55:05.7 &                &       & 15036.8 & 79.8 &    \\
256 & 17:02:31.3 & 78:43:05.4 & 16.19$\pm$0.08 & 16.68 & 15419.0 & 60.0 & 64 \\
262 & 17:02:34.2 & 78:39:37.6 & 16.25$\pm$0.07 &       & 17800.3 & 30.7 &    \\
265 & 17:02:34.6 & 78:31:28.9 & 17.56$\pm$0.08 &       & 16665.0 & 65.4 &    \\
266 & 17:02:35.7 & 78:48:12.5 &                &       & 16828.9 & 36.6 &    \\
267 & 17:02:36.0 & 78:45:58.3 & 16.04$\pm$0.08 &       & 18918.0 & 35.6 &    \\
270 & 17:02:37.0 & 78:40:36.0 & 16.63$\pm$0.07 &       & 17779.4 & 37.0 &    \\
272 & 17:02:38.1 & 78:33:34.3 & 16.05$\pm$0.08 & 16.60 & 15453.2 & 44.8 & 65 \\
294 & 17:02:46.3 & 78:51:46.7 &                & 15.52 & 16573.7 & 52.6 & 69 \\
295 & 17:02:46.8 & 78:25:46.7 &                &       & 53624.7 & 114.0&    \\
297 & 17:02:47.5 & 78:37:17.7 & 16.85$\pm$0.07 &       & 62676.0 & 109.0&    \\
298 & 17:02:47.6 & 78:44:27.9 & 14.58$\pm$0.08 & 14.45 & 20119.5 & 54.1 & 68 \\
301 & 17:02:48.9 & 78:38:29.6 & 15.86$\pm$0.07 & 15.92 & 16936.0 & 39.0 & 67 \\
310 & 17:02:55.2 & 78:35:04.5 & 16.41$\pm$0.07 &       & 16356.5 & 109.0&    \\
320 & 17:02:57.7 & 78:42:06.3 & 17.40$\pm$0.08 &       & 15976.8 & 67.7 &    \\
324 & 17:02:58.3 & 78:38:17.1 & 17.20$\pm$0.07 &       & 15763.1 & 34.4 &    \\
325 & 17:02:58.5 & 78:49:47.7 &                &       & 17654.6 & 47.1 &    \\
332 & 17:03:01.1 & 78:36:52.5 & 15.76$\pm$0.07 & 16.16 & 16545.0 & 35.0 & 71 \\
335 & 17:03:02.4 & 78:35:56.1 & 15.06$\pm$0.07 & 15.05 & 16303.0 & 100.0& 72 \\
347 & 17:03:08.8 & 78:41:21.0 & 16.11$\pm$0.07 & 16.74 & 14280.8 & 49.3 & 74 \\
351 & 17:03:10.4 & 78:14:40.9 &                & 15.89 & 17973.4 & 48.3 & 70 \\
353 & 17:03:11.3 & 78:40:31.1 & 15.80$\pm$0.07 & 16.31 & 15901.0 & 39.0 & 75 \\
358 & 17:03:12.0 & 78:33:09.3 & 16.01$\pm$0.05 & 16.42 & 16504.2 & 38.4 & 73 \\
360 & 17:03:12.2 & 78:40:21.4 & 17.88$\pm$0.07 & 17.80 & 16637.0 & 52.0 & 76 \\
364 & 17:03:13.0 & 78:45:56.4 & 16.24$\pm$0.08 & 16.65 & 15660.0 & 39.0 & 77 \\
369 & 17:03:16.9 & 78:26:46.4 &                &       & 16202.6 & 33.4 &    \\
383 & 17:03:21.5 & 78:43:54.1 & 16.48$\pm$0.08 &       & 15060.2 & 66.0 &    \\
385 & 17:03:22.1 & 78:46:56.2 & 16.18$\pm$0.08 &       & 19723.0 & 44.7 &    \\
394 & 17:03:24.8 & 78:20:39.1 &                &       & 18886.0 & 47.9 &    \\
396 & 17:03:25.3 & 78:35:30.8 & 17.53$\pm$0.07 &       & 16018.4 & 94.2 &    \\
399 & 17:03:26.0 & 78:52:42.7 &                &       & 19105.9 & 40.5 &    \\
407 & 17:03:28.3 & 78:36:51.5 & 16.21$\pm$0.07 & 16.71 & 17901.8 & 59.3 & 78 \\
410 & 17:03:29.2 & 78:37:55.7 & 15.02$\pm$0.07 & 14.00 & 17796.4 & 43.5 & 79 \\
414 & 17:03:29.7 & 78:39:55.1 & 15.07$\pm$0.07 & 15.30 & 17558.0 & 36.0 & 80 \\
417 & 17:03:32.5 & 78:37:02.9 & 17.58$\pm$0.07 &       & 16798.6 & 45.0 &    \\
420 & 17:03:33.8 & 78:37:48.3 & 14.54$\pm$0.07 & 14.20 & 16928.9 & 47.0 & 81 \\
422 & 17:03:34.1 & 78:50:55.2 &                & 16.28 & 18943.2 & 49.2 & 85 \\
426 & 17:03:35.3 & 78:39:54.6 & 16.09$\pm$0.07 & 16.57 & 15425.6 & 43.2 & 82 \\
428 & 17:03:35.6 & 78:37:45.1 & 14.34$\pm$0.07 & 14.00 & 15830.0 & 100.0& 83 \\
431 & 17:03:36.0 & 78:47:44.0 &                & 14.84 & 19800.0 & 100.0& 86 \\
436 & 17:03:36.9 & 78:40:31.1 & 17.26$\pm$0.07 &       & 16980.6 & 35.1 &    \\
438 & 17:03:37.5 & 78:38:27.7 & 16.15$\pm$0.07 & 16.63 & 17628.0 & 46.0 & 84 \\
439 & 17:03:37.8 & 78:46:58.6 & 17.23$\pm$0.08 &       & 19856.1 & 101.0&    \\
440 & 17:03:37.9 & 78:36:54.8 & 16.95$\pm$0.07 &       & 15848.1 & 74.2 &    \\
444 & 17:03:39.1 & 78:22:18.7 &                &       & 53949.6 & 84.9 &    \\
448 & 17:03:41.9 & 78:39:16.7 & 18.51$\pm$0.07 &       & 14742.2 & 44.3 &    \\
456 & 17:03:46.3 & 78:37:24.4 & 17.39$\pm$0.07 &       & 17885.8 & 37.7 &    \\
458 & 17:03:46.8 & 78:34:35.9 & 16.72$\pm$0.05 &       & 17323.9 & 40.2 &    \\
470 & 17:03:50.3 & 78:37:02.9 & 16.30$\pm$0.07 & 16.75 & 20245.0 & 41.9 & 88 \\
477 & 17:03:53.3 & 78:29:11.4 & 17.48$\pm$0.05 &       & 16265.6 & 52.9 &    \\
484 & 17:03:55.5 & 78:44:20.5 & 17.51$\pm$0.08 &       & 15648.7 & 72.2 &    \\
486 & 17:03:56.2 & 78:44:43.9 & 15.59$\pm$0.08 & 15.80 & 17115.2 & 36.2 & 91 \\
487 & 17:03:56.3 & 78:37:36.3 & 15.76$\pm$0.07 & 16.16 & 15151.1 & 34.5 & 89 \\
495 & 17:03:57.6 & 78:41:26.4 & 16.10$\pm$0.08 & 16.50 & 20404.0 & 46.0 & 90 \\
498 & 17:03:58.4 & 78:32:34.7 & 16.67$\pm$0.05 &       & 17010.7 & 53.9 &    \\
512 & 17:04:03.8 & 78:33:23.2 & 16.28$\pm$0.05 &       & 17355.3 & 53.1 &    \\
516 & 17:04:05.6 & 78:31:50.7 & 17.34$\pm$0.05 &       & 16181.8 & 39.8 &    \\
520 & 17:04:08.0 & 78:26:47.2 &                &       & 16803.9 & 53.1 &    \\
526 & 17:04:11.8 & 78:34:49.0 & 16.87$\pm$0.05 &       & 19642.6 & 38.5 &    \\
530 & 17:04:13.1 & 78:37:44.3 & 14.20$\pm$0.07 & 13.92 & 16903.0 & 100.0& 93 \\
538 & 17:04:14.3 & 78:54:35.1 &                &       & 18981.0 & 44.7 &    \\
543 & 17:04:16.1 & 78:40:45.7 & 15.48$\pm$0.07 & 15.99 & 15690.0 & 72.0 & 94 \\
547 & 17:04:17.0 & 78:47:32.3 &                &       & 19233.4 & 39.4 &    \\
565 & 17:04:20.9 & 78:36:22.3 & 17.79$\pm$0.07 &       & 17409.3 & 71.9 &    \\
568 & 17:04:21.6 & 78:42:30.5 & 16.42$\pm$0.08 &       & 14994.1 & 79.9 &    \\
581 & 17:04:26.8 & 78:38:26.1 & 14.18$\pm$0.07 & 13.34 & 17797.5 & 40.0 & 95 \\
585 & 17:04:27.9 & 78:25:19.1 &                &       & 16966.3 & 30.2 &    \\
586 & 17:04:28.1 & 78:45:28.3 & 18.20$\pm$0.08 &       & 17533.3 & 135. &    \\
591 & 17:04:29.7 & 78:38:48.0 & 17.04$\pm$0.07 &       & 16752.0 & 77.4 &    \\
597 & 17:04:32.6 & 78:42:20.4 & 16.74$\pm$0.08 &       & 15188.8 & 38.3 &    \\
605 & 17:04:35.6 & 78:40:05.9 & 17.34$\pm$0.07 &       & 19700.9 & 40.0 &    \\
606 & 17:04:35.9 & 78:22:47.2 &                &       & 17492.8 & 34.3 &    \\
607 & 17:04:37.0 & 78:53:21.1 &                &       & 18122.8 & 40.8 &    \\
616 & 17:04:40.2 & 78:41:31.2 & 17.63$\pm$0.08 &       & 18596.9 & 128.0&    \\
617 & 17:04:40.2 & 78:37:42.0 & 16.61$\pm$0.07 &       & 19674.9 & 103.0&    \\
620 & 17:04:40.6 & 78:13:49.5 &                & 16.41 & 18008.1 & 56.1 & 96 \\
621 & 17:04:41.4 & 78:32:09.1 & 16.75$\pm$0.05 &       & 16717.9 & 33.8 &    \\
623 & 17:04:43.4 & 78:37:22.8 & 15.58$\pm$0.07 & 16.07 & 15105.5 & 42.6 & 97 \\
633 & 17:04:47.1 & 78:40:31.5 & 17.01$\pm$0.07 &       & 16600.8 & 86.3 &    \\
634 & 17:04:47.1 & 78:34:33.9 & 16.34$\pm$0.05 &       & 18816.7 & 42.0 &    \\
635 & 17:04:47.7 & 78:38:30.0 & 14.76$\pm$0.07 & 14.76 & 19094.2 & 49.1 & 98 \\
637 & 17:04:48.3 & 78:48:21.3 &                & 15.44 & 19663.1 & 46.9 &100 \\
641 & 17:04:48.9 & 78:37:43.3 & 17.44$\pm$0.07 &       & 17215.2 & 37.8 &    \\
647 & 17:04:51.5 & 78:40:27.3 & 16.64$\pm$0.07 &       & 17734.3 & 63.9 &    \\
651 & 17:04:53.6 & 78:29:54.3 & 16.98$\pm$0.05 &       & 52295.4 & 114.0&    \\
656 & 17:04:55.3 & 78:51:07.2 &                &       & 19980.1 & 44.8 &    \\
658 & 17:04:55.8 & 78:34:04.2 & 17.75$\pm$0.05 &       & 18478.0 & 45.1 &    \\
659 & 17:04:56.3 & 78:43:06.4 & 17.09$\pm$0.08 &       & 18682.0 & 44.9 &    \\
662 & 17:04:57.4 & 78:38:13.5 & 17.25$\pm$0.07 &       & 19525.8 & 76.5 &    \\
664 & 17:04:57.7 & 78:50:04.8 &                & 16.27 & 19590.0 & 52.0 & 104\\
665 & 17:04:58.2 & 78:42:04.2 & 16.61$\pm$0.08 &       & 17351.7 & 37.8 &    \\
667 & 17:04:58.7 & 78:24:57.6 &                & 16.49 & 26841.1 & 49.0 &  99\\
668 & 17:04:58.8 & 78:35:05.1 & 16.23$\pm$0.07 & 16.77 & 17135.9 & 44.4 & 102\\
678 & 17:05:03.0 & 78:40:03.2 & 16.01$\pm$0.07 & 16.40 & 16762.0 & 49.0 & 105\\
684 & 17:05:06.3 & 78:28:22.8 &                & 15.65 & 19248.6 & 57.8 & 103\\
693 & 17:05:10.0 & 78:31:36.0 & 15.86$\pm$0.05 & 16.36 & 16535.7 & 34.8 & 106\\
697 & 17:05:11.7 & 78:48:14.9 &                & 16.72 & 19942.3 & 47.8 & 107\\
702 & 17:05:13.8 & 78:29:19.5 & 16.88$\pm$0.05 &       & 16484.4 & 37.3 &    \\
708 & 17:05:15.2 & 78:29:51.1 & 18.73$\pm$0.05 &       & 18146.7 & 39.8 &    \\
709 & 17:05:15.2 & 78:48:12.8 &                &       & 42213.8 & 54.4 &    \\
710 & 17:05:15.5 & 78:30:02.6 & 16.95$\pm$0.05 &       & 17939.0 & 73.7 &    \\
712 & 17:05:15.8 & 78:32:37.5 & 16.90$\pm$0.05 &       & 18122.4 & 106.0&    \\
715 & 17:05:18.1 & 78:38:46.1 & 16.42$\pm$0.07 & 16.55 & 17682.0 & 47.0 & 108\\
725 & 17:05:20.7 & 78:47:20.9 & 17.85$\pm$0.08 &       & 17771.8 & 44.9 &    \\
728 & 17:05:21.1 & 79:08:04.9 &                &       & 18806.5 & 48.3 &    \\
731 & 17:05:23.7 & 78:40:23.8 & 17.38$\pm$0.07 &       & 18127.4 & 94.8 &    \\
734 & 17:05:25.0 & 78:21:59.7 &                &       & 17321.5 & 36.2 &    \\
738 & 17:05:26.9 & 78:38:47.8 & 16.45$\pm$0.07 & 16.25 & 19599.0 & 57.0 & 109\\
744 & 17:05:30.3 & 78:40:19.7 & 16.22$\pm$0.07 & 16.20 & 17194.0 & 65.0 & 110\\
745 & 17:05:30.5 & 78:55:15.3 &                &       & 17193.4 & 53.1 &    \\
746 & 17:05:31.5 & 78:35:50.6 & 18.70$\pm$0.07 &       & 18652.4 & 44.2 &    \\
747 & 17:05:32.4 & 78:39:02.1 & 16.83$\pm$0.07 &       & 15428.5 & 39.6 &    \\
751 & 17:05:33.1 & 78:41:42.1 & 18.24$\pm$0.08 &       & 17789.6 & 41.1 &    \\
752 & 17:05:33.1 & 78:51:17.1 &                & 15.98 & 19478.6 & 53.7 & 113\\
753 & 17:05:33.5 & 78:35:12.7 & 15.33$\pm$0.07 & 15.27 & 18791.3 & 50.7 & 111\\
755 & 17:05:34.2 & 78:30:34.2 & 17.56$\pm$0.05 &       & 19044.3 & 45.6 &    \\
770 & 17:05:39.1 & 78:37:31.5 & 16.05$\pm$0.07 & 15.59 & 16142.1 & 34.7 & 112\\
780 & 17:05:42.2 & 78:36:25.4 & 17.39$\pm$0.07 &       & 17538.3 & 78.2 &    \\
789 & 17:05:46.9 & 78:29:12.0 & 18.06$\pm$0.05 &       & 39704.7 & 48.0 &    \\
792 & 17:05:47.6 & 78:39:57.5 &                &       & 18616.1 & 33.8 &    \\
795 & 17:05:48.6 & 78:22:34.0 &                & 15.94 & 16437.0 & 45.0 & 114\\
798 & 17:05:49.3 & 78:16:18.4 &                &       & 51621.4 & 155.0&    \\
799 & 17:05:49.5 & 78:31:07.5 & 17.47$\pm$0.05 &       & 17701.6 & 135.0&    \\
801 & 17:05:50.2 & 78:22:06.0 &                &       & 52074.8 & 65.0 &    \\
822 & 17:06:00.1 & 79:06:43.2 &                & 15.74 & 17927.3 & 47.4 & 117\\
847 & 17:06:11.1 & 78:24:38.2 &                &       & 18317.7 & 47.9 &    \\
848 & 17:06:11.4 & 78:32:50.4 & 15.14$\pm$0.05 & 15.42 & 18401.7 & 38.4 & 115\\
853 & 17:06:14.2 & 78:33:58.1 & 17.77$\pm$0.05 &       & 17388.7 & 38.2 &    \\
855 & 17:06:15.4 & 78:26:53.4 &                &       & 32062.0 & 84.6 &    \\
877 & 17:06:22.5 & 78:34:20.7 & 17.01$\pm$0.05 &       & 16658.2 & 107.0&    \\
882 & 17:06:24.0 & 79:02:37.3 &                &       & 17585.6 & 37.9 &    \\
889 & 17:06:28.2 & 78:49:13.9 &                &       & 19093.1 & 44.6 &    \\
895 & 17:06:31.3 & 78:25:18.0 &                & 15.61 & 15626.0 & 35.0 & 118\\
897 & 17:06:31.8 & 78:42:08.9 & 18.09$\pm$0.08 &       & 18858.9 & 27.8 &    \\
898 & 17:06:32.2 & 78:25:00.7 &                &       & 17957.7 & 61.8 &    \\
902 & 17:06:34.9 & 78:21:55.4 &                &       & 16799.8 & 43.3 &    \\
903 & 17:06:35.1 & 78:26:00.2 &                &       & 27178.6 & 103.0&    \\
917 & 17:06:39.7 & 78:25:57.3 &                &       & 52150.3 & 92.5 &    \\
921 & 17:06:42.9 & 78:35:58.3 & 16.27$\pm$0.07 & 16.72 & 14909.8 & 77.2 & 120\\
922 & 17:06:43.0 & 78:35:19.0 & 20.04$\pm$0.08 &       & 17954.2 & 154.0&    \\
934 & 17:06:48.6 & 78:35:33.8 &                &       & 17408.7 & 76.3 &    \\
935 & 17:06:49.5 & 78:28:40.9 & 17.30$\pm$0.05 &       & 20079.3 & 56.8 &    \\
939 & 17:06:51.6 & 78:33:42.5 & 16.42$\pm$0.05 &       & 18457.4 & 57.8 &    \\
943 & 17:06:53.3 & 78:30:56.1 & 15.63$\pm$0.05 & 16.15 & 17998.0 & 61.0 & 121\\
946 & 17:06:56.0 & 78:41:10.3 & 15.75$\pm$0.07 & 15.60 & 16866.5 & 37.6 & 122\\
948 & 17:06:56.4 & 78:15:14.4 &                &       & 16574.0 & 58.2 &    \\
950 & 17:06:56.6 & 78:23:18.0 &                &       & 52117.8 & 99.6 &    \\
956 & 17:07:00.3 & 78:41:24.3 & 17.38$\pm$0.07 &       & 17160.6 & 79.1 &    \\
962 & 17:07:01.7 & 79:10:19.6 &                & 16.43 & 11436.4 & 49.2 & 124\\
969 & 17:07:04.3 & 78:35:57.7 & 17.49$\pm$0.07 &       & 15917.3 & 32.7 &    \\
973 & 17:07:07.4 & 78:38:34.6 & 15.43$\pm$0.07 & 15.80 & 18843.9 & 37.8 & 123\\
980 & 17:07:11.5 & 78:38:58.4 & 17.15$\pm$0.07 &       & 17897.0 & 36.8 &    \\
984 & 17:07:13.4 & 78:53:18.9 &                & 15.93 & 16523.5 & 42.2 & 125\\
987 & 17:07:14.2 & 78:19:12.4 &                &       & 17609.7 & 79.7 &    \\
989 & 17:07:15.5 & 78:30:08.4 & 16.37$\pm$0.05 &       & 18581.5 & 47.5 &    \\
999 & 17:07:19.1 & 78:39:50.5 &                &       & 18473.7 & 37.8 &    \\
1002& 17:07:20.9 & 78:53:37.4 &                &       & 16563.9 & 34.9 &    \\
1004& 17:07:20.9 & 78:24:09.2 &                &       & 51293.1 & 108.0&    \\
1008& 17:07:26.6 & 78:36:22.0 &                & 16.47 & 17553.7 & 45.9 & 126\\
1009& 17:07:27.7 & 78:36:40.2 &                & 16.32 & 17637.6 & 47.2 & 127\\
1010& 17:07:28.8 & 78:30:08.4 &                &       & 19040.4 & 36.5 &    \\
1011& 17:07:30.4 & 78:24:41.8 &                &       & 51597.8 & 82.1 &    \\
1012& 17:07:31.4 & 79:08:36.8 &                & 15.49 & 16603.7 & 38.6 & 128\\
1013& 17:07:39.6 & 78:56:56.0 &                &       & 53835.5 & 61.0 &    \\
1014& 17:07:54.9 & 78:44:48.8 &                &       & 17027.4 & 38.2 &    \\
1015& 17:07:58.5 & 78:18:39.6 &                &       & 18428.3 & 42.0 &    \\
1016& 17:08:02.0 & 78:17:34.7 &                &       & 39920.8 & 78.6 &    \\
1017& 17:08:02.9 & 78:32:38.3 &                &       & 17910.5 & 34.2 &    \\
1018& 17:08:05.4 & 78:37:32.0 &                & 16.29 & 17841.0 & 49.0 & 129\\
1019& 17:08:05.7 & 78:22:57.1 &                &       & 52032.8 & 81.9 &    \\
1020& 17:08:09.2 & 78:34:16.7 &                &       & 18541.1 & 47.1 &    \\
1021& 17:08:10.6 & 78:39:13.1 &                &       & 15020.8 & 72.5 &    \\
1022& 17:08:13.9 & 78:46:48.4 &                &       & 15957.6 & 53.6 &    \\
1023& 17:08:22.9 & 78:22:28.1 &                &       & 17243.5 & 29.9 &    \\
1024& 17:08:24.5 & 78:34:28.8 &                &       & 18009.2 & 79.8 &    \\
1025& 17:08:31.6 & 78:21:37.6 &                & 16.02 & 21638.7 & 52.3 & 130\\
1026& 17:08:33.1 & 78:29:54.4 &                & 15.40 & 18684.4 & 46.3 & 131\\
1027& 17:08:35.8 & 78:40:31.6 &                &       & 17215.9 & 29.8 &    \\
1028& 17:08:41.3 & 78:23:46.1 &                &       & 17017.6 & 65.2 &    \\
1029& 17:08:45.6 & 78:34:03.4 &                &       & 18347.9 & 34.3 &    \\
1030& 17:08:46.3 & 78:37:03.1 &                &       & 16512.8 & 37.3 &    \\
1031& 17:08:47.5 & 78:25:55.8 &                &       & 16860.2 & 6.85 &    \\
1033& 17:08:57.7 & 78:50:15.3 &                & 15.57 & 16064.1 & 50.9 & 134\\
1034& 17:08:58.7 & 78:32:58.7 &                & 16.75 & 16191.4 & 17.4 & 133\\
1035& 17:08:59.0 & 78:31:40.0 &                &       & 51442.5 & 112.0&    \\
1036& 17:09:00.6 & 78:46:19.4 &                & 15.46 & 11843.8 & 26.2 & 135\\
1037& 17:09:09.7 & 78:52:32.0 &                & 15.55 & 18811.8 & 46.3 & 137\\
1039& 17:09:12.7 & 78:50:56.8 &                &       & 52250.8 & 120.0&    \\
1040& 17:09:14.6 & 78:38:10.8 &                &       & 18134.8 & 32.7 &    \\
1041& 17:09:15.1 & 78:21:58.4 &                &       & 16702.5 & 50.5 &    \\
1042& 17:09:16.2 & 78:27:11.6 &                &       & 7427.43 & 48.6 &    \\
1043& 17:09:18.7 & 78:30:16.8 &                &       & 16700.3 & 43.9 &    \\
1044& 17:09:22.4 & 78:22:03.4 &                &       & 16193.1 & 63.5 &    \\
1045& 17:09:23.9 & 78:42:14.6 &                & 16.39 & 17927.9 & 29.0 & 138\\
1046& 17:09:30.2 & 78:45:46.3 &                &       & 18249.8 & 44.8 &    \\
1047& 17:09:39.5 & 78:33:07.9 &                & 15.73 & 16119.9 & 32.5 & 142\\
1048& 17:09:43.2 & 78:31:03.8 &                & 16.63 & 18561.0 & 38.0 & 141\\
1050& 17:09:44.9 & 78:20:18.0 &                & 15.12 & 17077.0 & 63.2 & 140\\
1051& 17:09:45.8 & 78:33:06.7 &                &       & 18694.1 & 58.3 &    \\
1052& 17:09:46.6 & 78:41:33.2 &                & 16.58 & 17892.0 & 42.0 & 143\\
1053& 17:10:07.0 & 78:36:56.6 &                &       & 59089.3 & 51.8 &    \\
1054& 17:10:13.4 & 78:39:22.1 &                & 16.44 & 17043.0 & 53.0 & 144\\
1055& 17:10:27.4 & 78:22:47.8 &                & 16.70 & 17045.5 & 36.3 & 145\\
1056& 17:10:31.2 & 78:29:10.3 &                & 16.54 & 15791.5 & 65.5 & 146\\
1057& 17:10:31.6 & 78:33:51.3 &                & 16.68 & 19464.0 & 34.1 & 147\\
1058& 17:10:45.9 & 78:32:15.3 &                & 15.11 & 18767.0 & 31.0 & 148\\
1059& 17:10:58.8 & 78:45:26.1 &                &       & 27118.1 & 59.4 &    \\
1060& 17:10:59.3 & 78:36:52.4 &                &       & 41643.9 & 36.8 &    \\
1061& 17:11:20.4 & 78:49:47.7 &                &       & 39492.5 & 83.9 &    \\
1062& 17:11:49.5 & 78:35:52.7 &                &       & 17819.7 & 101.0&    \\
1065& 17:11:58.3 & 78:36:26.8 &                &       & 42133.1 & 108.0&    \\
1068& 17:12:37.1 & 78:28:30.9 &                & 16.33 & 17125.5 & 67.1 & 153\\
1069& 17:12:51.2 & 79:03:39.6 &                & 16.37 & 21886.4 & 45.4 & 156\\
1071& 17:12:57.1 & 78:56:44.1 &                &       & 19978.3 & 44.7 &    \\
1072& 17:13:01.1 & 78:29:02.8 &                & 15.66 & 17540.5 & 50.0 & 155\\
1073& 17:13:04.8 & 78:34:24.2 &                &       & 19111.7 & 79.2 &    \\
1074& 17:13:08.5 & 78:33:41.7 &                & 15.90 & 17432.0 & 52.2 & 157\\
1075& 17:13:08.6 & 78:33:28.0 &                & 14.95 & 15928.6 & 76.4 & 158\\
1078& 17:13:54.0 & 78:46:29.9 &                &       & 25936.1 & 79.1 &    \\
1080& 17:13:56.8 & 78:59:33.5 &                & 14.87 & 19613.7 & 61.5 & 164\\
\enddata
\end{deluxetable}

\begin{deluxetable}{rll}
\tablewidth{0pt}
\tablecaption{Statistical Tests\label{TESTS}}
\tablerefs{
[1]--(Beers, Flynn \& Gebhardt 1990);
[2]--(Bird \& Beers 1993);
[3]--(Bird 1994);
[4]--(D'Agostino \& Stephens 1986);
[5]--(Dressler \& Shectman 1988);
[6]--(Fitchett 1988);
[7]--(Heisler, Tremaine \& Bahcall 1985);
[8]--(Iglewicz 1983);
[9]--(Lee 1979);
[10]--(Pearson \& Hartley 1962);
[11]--(Pinkney et al.\ 1996);
[12]--(Rhee, van Haarlem \& Katgert 1991);
[13]--(Shapiro \& Wilk 1965);
[14]--(Stephens 1986);
[15]--(West \& Bothun 1990);
[16]--(Yahil \& Vidal 1977)}

\tablehead{
	\colhead{Test}	&
	\colhead{Description}	&
	\colhead{References}	
}
\startdata
a	& Average absolute deviation from mean velocity & [1], [10], [16]\\
u	& Maximum velocity separation & [1], [16]\\
W	& Variance of pairwise velocity separations & [1], [13], [16] \\
$\sqrt{b_1}$	& Skewness of distribution & [1], [4]\\
$b_2$	& Kurtosis of distribution & [1], [4], [16]\\
$\sqrt{b_1}b_2$ Omnibus & Skewness and Kurtosis of distribution &
	[1], [4]\\
I	& Ratio of standard deviation with median estimators & [1], [8] \\
	& and standard deviation with biweight estimators & \\
KS	& Deviation of CDF from best fit Gaussian\tablenotemark{a}
	& [1], [4]\\
V	& Deviation of CDF from best fit Gaussian
	& [1], [4]\\
$W^{2}$	& Deviation of CDF from best fit Gaussian & [1], [4]\\
$U^{2}$	& Deviation of CDF from best fit Gaussian & [1], [4]\\
$A^{2}$	& Deviation of CDF from best fit Gaussian & [1], [4]\\
Lee 2D 	& Relative values of spaces and pairs in two dimensions & [6],
	[9], [11]\\ 
Fourier Elongation	& Elongation of spatial distribution &[11], [12] \\ 
\\
\\
\\   
\\
\\
$\Delta$	& Deviation of local mean and sigma from global values
	& [5]\\ 
$\epsilon$	& Change of projected mass estimator as a function & [3], [7] \\
	& of positional space  & \\
$\alpha$	& Shift of spatial centroid as a function of velocity space
	window & [11], [15]\\
Lee 3D	& Relative values of spaces and pairs in three dimensions &
	[6], [9], [11]\\ 
\enddata
\tablenotetext{a}{\underline{c}umulative \underline{d}istribution \underline{f}unction}
\end{deluxetable}

\begin{deluxetable}{ccc}
\tablewidth{0pt}
\tablecaption{1-D Test results\label{ROSTAT}}
\tablehead{
	\colhead{Statistical}	&
	\colhead{Value}	&
	\colhead{Significance}	\\
	\colhead{Test}	&
	\colhead{}	&
	\colhead{}	
}
\startdata
  a		&		0.82	& \bf {0.05}	\\
  u		&		4.68	& \bf {$<$0.05}	\\
  W		&		0.97	& \bf {$<$0.01} \\
  $\sqrt{b_1}$		&		$-$0.011	&	0.47\\
  $b_2$		&		2.36	& \bf {$<$0.01} \\
  $\sqrt{b_1}b_2$ Omnibus	&		9.174	&\bf {0.01}\\
  I		&		0.94	&	$>$0.10\\
  KS		&		0.50	&	0.25\\
  V		&		0.98	&	0.25\\
  $W^{2}$		&		0.05	&	0.49\\
  $U^{2}$		&		0.05	&	0.44\\
  $A^{2}$		&		0.45	&	0.27\\
\enddata
\end{deluxetable}

\begin{deluxetable}{ccc}
\tablewidth{0pt}
\tablecaption{2-D Test results\label{2dTESTS}}
\tablehead{
	\colhead{Statistical}	&
	\colhead{Value}	&
	\colhead{Significance}	\\
	\colhead{Test}	&
	\colhead{}	&
	\colhead{}	
}
\startdata
Lee 2D $l_{max}$	&	1.58	&	\bf{0.01} \\
Lee 2D $l_{rat}$	&	1.29	&	\bf{$<$0.01} \\
Fourier-Elongation	&	0.92	&	0.30 \\

\enddata
\end{deluxetable}

\begin{deluxetable}{ccc}
\tablewidth{0pt}
\tablecaption{3-D Test results\label{3DTESTS}}
\tablehead{
	\colhead{Statistical}	&
	\colhead{Value}	&
	\colhead{Significance}	\\
	\colhead{Test}	&
	\colhead{}	&
	\colhead{}	
}
\startdata
$\Delta$	&	374.1				&	\bf{$<$0.01} \\
$\epsilon$	&	$2.24 \times 10^{15} M_{\sun}$	&	0.68 \\
$\alpha$	&	0.02 Mpc			&	\bf{0.06} \\
Lee 3D $l_{max}$	&	2.29			&	0.22 \\
\enddata
\end{deluxetable}

\begin{deluxetable}{ccccccc}
\tablewidth{0pt}
\tablecaption{KMM Results\label{KMM3T}}
\tablehead{
	\colhead{}	&
	\colhead{} &
	\colhead{} &
	\colhead{Velocity}	&
	\colhead{Velocity}	&
	\colhead{}	&	
	\colhead{} \\
	\colhead{Gaussian}	&
	\colhead{RA} &
	\colhead{Dec} &
	\colhead{Mean}	&
	\colhead{Sigma}	&
	\colhead{$N_{gal}$}	&	
	\colhead{$M_{PME}$} \\
	\colhead{}	&
	\multicolumn{2}{c}{(J2000.0)}&
	\multicolumn{2}{c}{(\kmsec)}&
	\colhead{}	&	
	\colhead{($10^{15} M_{\odot}$)} 
}
\startdata
1 &17:01:57&78:39:37 & 15701.78	& 550.77	& 53 	& $0.51$\\
2 &17:04:28&78:38:35 & 17665.59	& 839.41	& 196 	& $1.6$\\
3 &17:03:52&78:46:08 & 19687.67	& 304.38	& 28 	& $0.17$\\
\enddata
\end{deluxetable}

\begin{deluxetable}{cccccccc}
\tablewidth{0pt}
\tablecaption{Brightest galaxies in A2256\label{BRIGHT}}
\tablehead{
	\colhead{Brightness}	&
	\colhead{}	&
	\colhead{Fabricant} &
	\colhead{}	&
	\colhead{}	&
	\colhead{} &
	\colhead{} &
	\colhead{} \\
	\colhead{Rank}	&
	\colhead{ID}	&
	\colhead{ID} &
	\colhead{RA} &
	\colhead{Dec} &
	\colhead{Group} &
	\colhead{$M_{R\KC}$} &
	\colhead{$V_{r}$} \\
	\colhead{}	&
	\colhead{}	&
	\colhead{} &
	\multicolumn{2}{c}{(J2000.0)}& 
	\colhead{} &
	\colhead{} &
	\colhead{(\kmsec)}
}
\startdata
1&581& 95 & 17:04:26.9 & 78:38:26.1 & 2 & 14.19$\pm$0.07 & 17797.6$\pm$40.0 \\
2&530& 93 & 17:04:13.2 & 78:37:44.3 & 2 & 14.20$\pm$0.07 & 16903.0$\pm$100.0 \\
3&428& 83 & 17:03:35.6 & 78:37:45.1 & 1 & 14.35$\pm$0.07 & 15830.0$\pm$100.0 \\
4&420& 81 & 17:03:33.8 & 78:37:48.3 & 2 & 14.55$\pm$0.07 & 16929.0$\pm$47.1 \\
5&298& 68 & 17:02:47.6 & 78:44:27.9 & 3 & 14.59$\pm$0.08 & 20119.5$\pm$54.2 \\
\enddata
\end{deluxetable}

\begin{deluxetable}{cccccc}
\tablecolumns{6}
\tablewidth{0pt}
\tablecaption{Radio Source Postions in A2256\label{RADIO}}
\tablehead{
	\colhead{Component} &
	\colhead{RA} &
	\colhead{Dec} &
	\colhead{ID}&
	\colhead{Group}&
	\colhead{$V_{r}$}\\
	\colhead{}&
	\multicolumn{2}{c}{(J2000.0)}& 
	\colhead{}&
	\colhead{}&
	\colhead{(\kmsec)}
}
\startdata
A  & 17:03:29.1 & 78:37:53.7 & 410 & 2 & 17796.4 \\
B  & 17:03:02.4 & 78:35:53.8 & 335 & 1 & 16303.0 \\
C  & 17:03:29.6 & 78:39:53.0 & 414 & 2 & 17558.0 \\
D  & 17:04:47.8 & 78:38:29.5 & 635 & 2 & 19094.2 \\
E  & 17:05:33.2 & 78:51:16.3 & 752 & 3 & 19478.6 \\
F1 & 17:02:17.9 & 78:41:28.3 & & & \\
F2 & 17:06:14.5 & 78:39:41.0 & & & \\
F3 & 17:06:51.5 & 78:41:06.3 & 946 & 2 & 16866.5 \\
G  & 17:03:43.0 & 78:50:03.3 & 486 & 2 & 17115.2 \\
H  & 17:02:31.3 & 78:42:37.8 & & & \\
I  & 17:00:51.8 & 78:41:19.8 & 88  & 2 & 17325.0 \\
J  & 17:01:11.9 & 78:43:26.5 & 116 & 1 & 16122.9 \\
K  & 17:02:18.3 & 78:46:02.2 & 233 & 3 & 19643.0 \\
L  & 17:02:50.3 & 78:31:48.4 & & & \\
M  & 17:09:29.3 & 78:41:38.6 & & & \\
N  & 17:01:46.4 & 78:00:00.1 & & & \\
O  & 17:05:14.4 & 78:30:17.8 & 706 & & \\
P  & 17:01:03.8 & 78:30:04.7 & 111 & & \\
Q  & 17:03:22.2 & 78:25:05.5 & & & \\
R  & 16:58:49.0 & 78:37:21.0 & & & \\
S  & 16:59:26.4 & 78:41:14.9 & & & \\
T  & 17:03:56.6 & 77:58:12.5 & & & \\
U  & 17:09:15.2 & 78:28:17.2 & & & \\
V  & 17:00:19.9 & 78:28:05.8 & & & \\
W  & 17:09:47.1 & 78:51:10.6 & & & \\
X  & 17:07:28.2 & 78:47:22.2 & & & \\
Y  & 17:09:18.5 & 78:29:02.6 & & & \\
Z  & 17:03:53.3 & 78:40:31.3 & & & \\
AA & 17:08:32.4 & 78:29:53.1 & & & \\
AB & 17:04:04.2 & 78:23:30.0 & & & \\
AC & 17:02:46.3 & 78:51:45.5 & & & \\
AD & 16:58:18.1 & 78:29:33.8 & & & \\
AE & 16:59:04.0 & 78:45:34.5 & & & \\
AF & 16:57:58.1 & 78:36:15.2 & & & \\
\enddata
\end{deluxetable}

\begin{figure}
\begin{center}
\leavevmode
\hbox{%
\epsfysize=5.0in
\epsffile{berrington.fig1.eps}}
\end{center}
\end{figure}

\begin{figure}
\begin{center}
\leavevmode
\hbox{%
\epsfysize=5.0in
\epsffile{berrington.fig2.eps}}
\end{center}
\end{figure}

\begin{figure}
\begin{center}
\leavevmode
\hbox{%
\epsfysize=5.0in
\epsffile{berrington.fig3.eps}}
\end{center}
\end{figure}

\begin{figure}
\begin{center}
\leavevmode
\hbox{%
\epsfysize=5.0in
\epsffile{berrington.fig4.eps}}
\end{center}
\end{figure}

\begin{figure}
\begin{center}
\leavevmode
\hbox{%
\epsfysize=5.0in
\epsffile{berrington.fig5.eps}}
\end{center}
\end{figure}

\begin{figure}
\begin{center}
\leavevmode
\hbox{%
\epsfysize=5.0in
\epsffile{berrington.fig6.eps}}
\end{center}
\end{figure}

\begin{figure}
\begin{center}
\leavevmode
\hbox{%
\epsfysize=5.0in
\epsffile{berrington.fig7.eps}}
\end{center}
\end{figure}

\begin{figure}
\begin{center}
\leavevmode
\hbox{%
\epsfysize=5.0in
\epsffile{berrington.fig8.eps}}
\end{center}
\end{figure}

\begin{figure}
\begin{center}
\leavevmode
\hbox{%
\epsfysize=5.0in
\epsffile{berrington.fig9.eps}}
\end{center}
\end{figure}

\begin{figure}
\begin{center}
\leavevmode
\hbox{%
\epsfysize=5.0in
\epsffile{berrington.fig10.eps}}
\end{center}
\end{figure}

\begin{figure}
\begin{center}
\leavevmode
\hbox{%
\epsfysize=5.0in
\epsffile{berrington.fig11.eps}}
\end{center}
\end{figure}

\begin{figure}
\begin{center}
\leavevmode
\hbox{%
\epsfysize=5.0in
}
\vspace{1.0in}
\end{center}
\end{figure}

\begin{figure}
\begin{center}
\leavevmode
\hbox{%
\epsfysize=5.0in
}
\vspace{1.0in}
\end{center}
\end{figure}

\end{document}